\documentclass[10pt, amssymb, aps, longbibliography, superscriptaddress, nofootinbib, floatfix,
]{revtex4-2}
\pdfoutput=1

\usepackage[table]{xcolor}
\usepackage{graphicx} 
\graphicspath{{figures/}}
\usepackage{comment}
\usepackage{bbold}
\usepackage{amsmath}
\usepackage{algorithm}
\usepackage{algpseudocodex}

\usepackage{braket}
\usepackage{tikz}
\usepackage{forest}
\usepackage{caption}
\usepackage{subcaption}
\usepackage{siunitx}
\usepackage{rotating}
\usepackage{tabularray}
\usepackage[compat=0.6]{yquant}
\useyquantlanguage{groups}
\usetikzlibrary{fit,quotes}
\usetikzlibrary{shapes.geometric,arrows,positioning}
\yquantset{plusctrl/.style={/yquant/every control/.style={/yquant/operators/every not}, every positive control/.style={}}}

\definecolor{riverlane_green}{RGB}{0, 111, 98}
\definecolor{riverlane_light_green}{RGB}{0, 150, 143}
\definecolor{riverlane_orange}{RGB}{255, 117, 0}
\definecolor{riverlane_red}{RGB}{220, 68, 5}
\definecolor{riverlane_pink}{RGB}{207, 111, 127}
\usepackage{hyperref}
\hypersetup{
  colorlinks   = true, 
  urlcolor     = riverlane_green, 
  linkcolor    = riverlane_orange, 
  citecolor   = riverlane_green  
}
\begin{document}
\title{Measurement Schemes for Quantum Linear Equation Solvers}
\author{Andrew Patterson}
\email{andrew.patterson@riverlane.com}
\affiliation{Riverlane, St.~Andrews House, 59 St.~Andrews Street, Cambridge CB2 3BZ, United Kingdom}
\author{Leigh Lapworth}
\email{leigh.lapworth@rolls-royce.com}
\affiliation{Rolls-Royce plc., P.O. Box 31, Derby, DE24 8BJ, United Kingdom}
\date{\today}

\begin{abstract}
  Solving Computational Fluid Dynamics (CFD) problems requires the inversion of a linear system of equations, which can be done using a quantum algorithm for matrix inversion~\cite{Gilyen2019}.
  However, the number of shots required to measure the output of the system can be prohibitive and remove any advantage obtained by quantum computing.
  In this work we propose a scheme for measuring the output of a Quantum Singular Value Transform (QSVT) matrix inversion algorithm specifically for the CFD use case.
  We use a Quantum Signal Processing (QSP) based amplitude estimation algorithm~\cite{Rall2023} and show how it can be combined with the QSVT matrix inversion algorithm.
  We perform a detailed resource estimation of the amount of computational resources required for a single iteration of amplitude estimation, and compare the costs of amplitude estimation with the cost of not doing amplitude estimation and measuring the whole wavefunction.
  We also propose a measurement scheme to reduce the number of amplitudes measured in the CFD example by focussing on large amplitudes only.
  We simulate the whole CFD loop, finding that thus measuring only a small number of the total amplitudes in the output vector still results in an acceptable level of overall error.
\end{abstract}

\maketitle

\section{Introduction}%
\label{sec:Introduction}

Solving systems of linear equations is one of the promising applications of quantum computers outside of chemistry simulations and the hidden subgroup problem for factoring \cite{Shor1994}.
In this paper we focus upon the measurement problem for a Quantum Linear Equations System (QLES) algorithm \cite{Lapworth2022b} based upon the Quantum Singular Value Transform (QSVT)~\cite{Gilyen2019}.
This algorithm uses matrix inversion to solve for $ \mathbf{x} $ the equation
\begin{equation}
  A \mathbf{x} = \mathbf{b},
  \label{eq:matrix_inversion}
\end{equation}
by inverting the matrix $ A $ and applying it to the input vector, $\mathbf{b}$.
At the completion of the algorithm all the qubits are measured and we post select upon a subset (the flag qubits) being measured all in the $\ket{0}$ state.
The probability of this outcome is affected by the subnormalisation, $\kappa$ of the matrix to be inverted, and can be quite small which requires many repeats of the circuit to measure the output.
Using an amplitude amplification algorithm is a common method of projecting onto the $\ket{0}^{\otimes_{\textrm{flag} }}$ state, and in this report we will detail a scheme that uses amplitude estimation to also project onto individual basis vectors representing elements of the corrections vector.
We will also use an amplitude estimation algorithm that is based on Quantum Signal Processing (QSP), similar to QSVT~\cite{Rall2023}.

\begin{figure}[ht]
  \begin{center}
    \begin{subfigure}[b]{0.45\textwidth}
      \includegraphics[width=\textwidth]{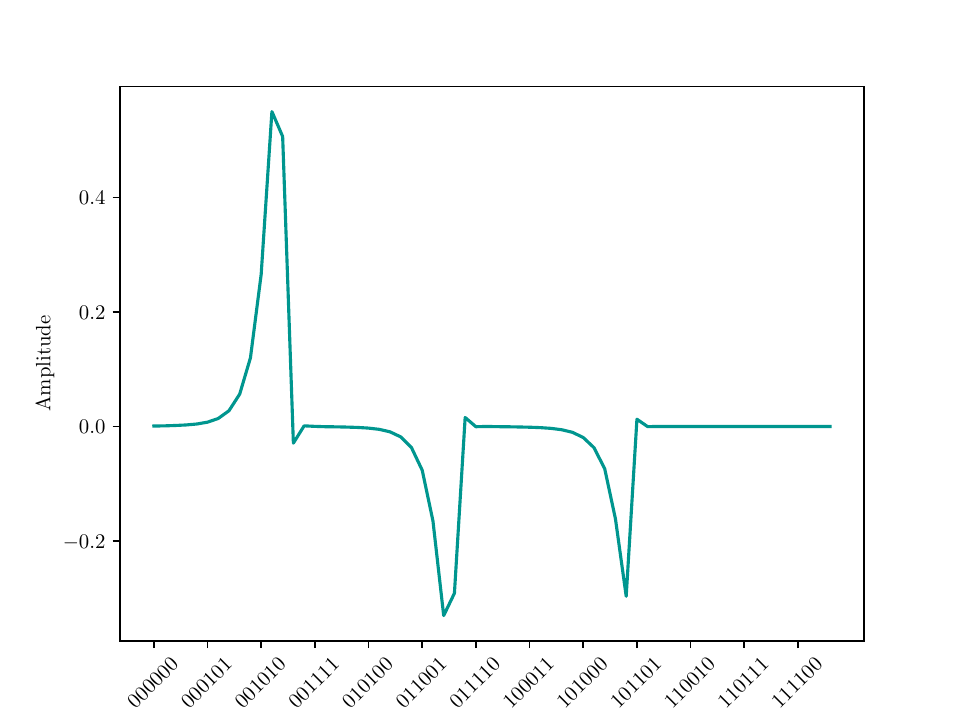}
        \caption{}\label{fig:output_vec_a}
    \end{subfigure}
    \begin{subfigure}[b]{0.45\textwidth}
      \includegraphics[width=\textwidth]{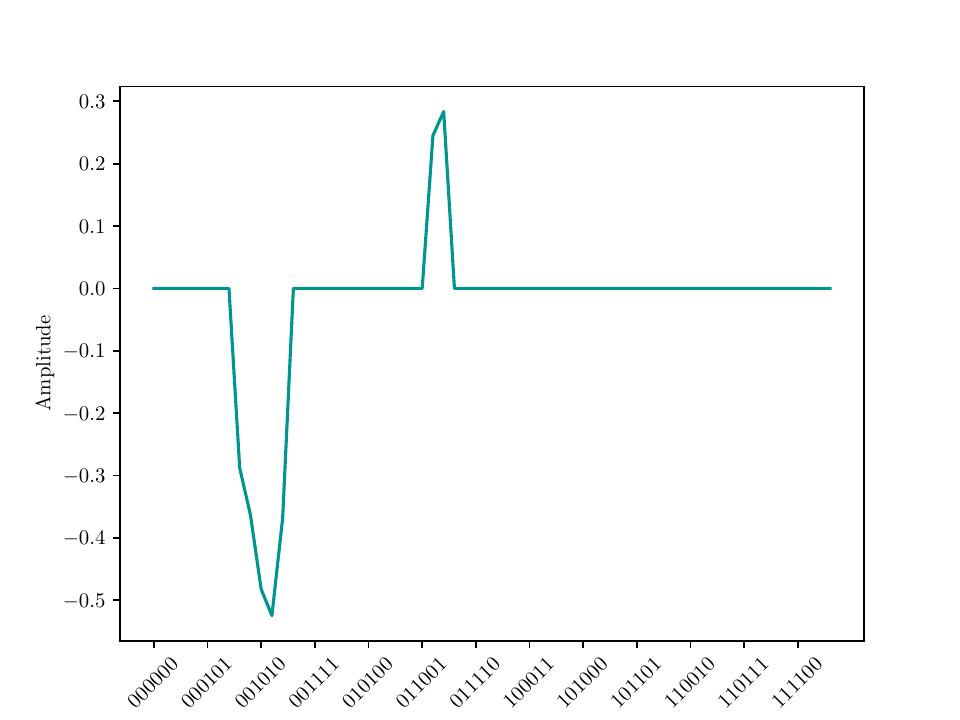}
        \caption{}\label{fig:output_vec_b}
    \end{subfigure}
  \end{center}
  \caption{Normalised output vectors of the QLES algorithm, these will be stored as probability amplitudes of the output qubits. ~\ref{fig:output_vec_a}: A vector from early in the iteration,  ~\ref{fig:output_vec_b}: A vector from late in the iteration.}\label{fig:output_vec}
\end{figure}

The QLES algorithm essentially uses a quantum algorithm (here QSVT) as a drop-in subroutine for the inversion of a matrix and return of a corrections vector.
The quantum computer solves the linear part of the problem, and we use existing classical methods to linearise the non-linear equations and apply returned corrections.
We are concerned here with measurement of the quantum state that describes this vector, which in the naive case requires $\frac{1}{\epsilon^2}$ measurements of the (post-selected) output qubits.
In the QLES algorithm $\mathbf{x}$ is a vector of corrections that we apply to the input vector in an iterative procedure until the system has converged.
We can use the nature of the corrections vector to make some simplifying assumptions.
Figure~\ref{fig:output_vec_a} shows an example output for a corrections vector from the beginning of a program, and~\ref{fig:output_vec_b} shows one from the final iteration.
We see that large peaks in the amplitude are present  at the beginning of the iteration, with a small number of larger peaks still being the dominant feature towards the end of the algorithm.
Using this feature we suggest that a measurement scheme based upon amplitude encoding is useful, where we estimate the amplitude of the largest peaks only and return this as an approximate correction.
As the corrections in Figure~\ref{fig:output_vec} also contain negative corrections, we must use a measurement scheme which is sensitive to the sign of the amplitude.

We denote the state at the end of the matrix inversion algorithm as $\ket{\psi}$, and the algorithm itself as implemented by the unitary $U_\psi$; this allows us to quantify the resources required to output a correction vector in the terms of oracle access to the circuit implementing $ U_\psi \ket{0^{\otimes n}} = \ket{\psi}$.
The naive approach to finding the corrections vector is to use quantum state tomography to accurately record the amplitude of each basis vector ($ \ket{00 \dots 0}, \ket{00 \dots 1} \dots $), which requires exponential calls to $ U_\psi $ in the number of qubits, $ n $~\cite{James2001}.
Advancements have been made in quantum state tomography, including maximum likelihood estimation (MLE) \cite{Hradil1997, James2001}, neural network based methods \cite{Torlai2018}, and tensor network based methods \cite{Cramer2010}.
If the state being measured can be represented as a Matrix Product State, MPS, measurement time is linear in $n$~\cite{Cramer2010}, neural network methods can be used to accurately reconstruct observables of the quantum state~\cite{Torlai2018}, but no guarantees on the number of shots required are given.
Alternatively, the quantum signal processing approach to Amplitude Estimation discussed in~\cite{Rall2023} requires a number of calls to the oracle that scales with the length of the polynomial used to construct the reflection operator.

In this paper we will present a measurement scheme using amplitude estimation from QSP.
The quantum algorithm combines ideas from~\cite{Rall2023} which shows how to do amplitude estimation using QSP and~\cite{Rossi2023a} which discusses embedding a QSVT algorithm within a QSP protocol. 
As the corrections vectors have a defined sign, we will use the real amplitude estimation scheme from~\cite{Manzano2023} to provide and estimate of the sign of the amplitude.
The combination of these three algorithms is discussed in Section~\ref{sec:amplitude_estimation_for_corrections_vectors}.
In Section~\ref{sec:measurement_scheme_for_corrections_vectors} we discuss the measurement scheme specifically for corrections vectors, where we propose reducing the number of amplitudes to measure, focussing on the peaks only.
We introduce the CFD scenario simulated in Section~\ref{sec:cfd_method}.
In Section~\ref{sec:results} we perform a resource estimation of measuring a single amplitude of the correction vector using amplitude estimation, and compare that cost to not running the amplitude estimation algorithm and receiving an estimator of all amplitudes, we also simulate the scheme discussed in Section~\ref{sec:measurement_scheme_for_corrections_vectors} within a classical CFD loop, comparing the performance of an algorithm with a high cut-off to that with a low cut-off (the threshold determining which amplitudes are considered).
Finally, in Section~\ref{sec:discussion} we will conclude and discuss the results.

\section{Amplitude Estimation for Corrections Vectors} 
\label{sec:amplitude_estimation_for_corrections_vectors}

In this section we will introduce the three algorithms we combine to measure the output vector of a matrix inversion problem using QSVT.
Firstly, we will review the algorithm of Rall et. al~\cite{Rall2023} where they show how to implement amplitude estimation using QSP, achieving a constant factor improvement and providing an estimate of query complexity to $Q_\Pi$, or the number of iterations of the amplitude estimation circuit.
As we are using a QSVT circuit to implement matrix inversion we need to show that the QSP protocol interacts with the matrix inversion algorithm correctly.
Secondly, we use the results of~\cite{Rossi2023a} to do this, using the fact that the polynomial approximating $f(A) \approx A^{-1}$ is real, and therefore an anti-symmetric list of phase factors can be generated.
Thirdly, as the amplitude estimation scheme as described so far can only estimate $|a|$ instead of $a$, we use the real amplitude estimation scheme of~\cite{Manzano2023} to modify our scheme to give the sign of $a$ also.
Finally, we combine the three algorithms discussed to give an overall query complexity for a single amplitude.

In~\cite{Rall2023} an amplitude estimation technique is developed that instead of using oracle access to reflections $Z_\Pi, Z_\psi$ about axes defined by a projector and a state, $\ket{\Pi}, \ket{\psi}$, assumes access to rotations around those axes, $e^{i\phi Z_\Pi}, e^{i\phi Z_\psi}$.
Access to the rotation operators allows us to use techniques from Quantum Signal Processing (QSP)~\cite{Gilyen2019} for a constant factor improvement in the number of oracle calls, and to use other useful techniques such as non-destructive amplitude estimation, where we return a copy of $\ket{\psi}$ at the end of the algorithm.
QSP techniques allow us to choose the rotation values $\phi$ such that we can implement a polynomial of the amplitude, $P(a)$.

\subsection{Vanilla Amplitude Estimation}
\label{sub:vanilla_amplitude_estimation} 

Amplitude estimation was first developed in Ref.~\cite{Brassard2002}, building upon a subroutine of Grover's algorithm~\cite{Grover2001} and Quantum Phase Estimation~\cite{Kitaev1996}.
In all amplitude estimation routines, we begin with the state we wish to investigate, $\ket{\psi}$, and a projector, $\Pi$, onto the subspace we wish to measure the amplitude, $a$, of, $ a = | \Pi \ket{\psi}|$.
The projector and its complement can be turned into basis vectors of a two-dimensional Hilbert space:
\begin{equation}
\begin{aligned}
  \ket{\Pi} &= a^{-1} \Pi \ket{\psi} \\
  \ket{\Pi^\perp} &= \bar{a}^{-1} (I - \Pi) \ket{\psi}
  \label{eq:projectors}
\end{aligned}
\end{equation}
where $a$ is the amplitude of the state we wish to measure,  and $\bar{a} = \sqrt{1 - a^2}$.
We can express $ \ket{\psi} $ in terms of this new orthonormal basis:
\begin{equation}
  \ket{\psi} = a \ket{\Pi} + \bar{a} \ket{\Pi^\perp},
  \label{eq:psi_projector}
\end{equation}
and define another basis vector to complement $\ket{\psi}$:
\begin{equation}
  \ket{\psi^\perp} = \bar{a} \ket{\Pi} - a  \ket{\Pi^\perp}.
  \label{eq:psi_perp}
\end{equation}
We use these two orthonormal bases to define the reflection operators, 
\begin{equation}
\begin{aligned}
  Z_\psi &= \ket{\psi}\bra{\psi} - \ket{\psi^\perp}\bra{\psi^\perp} \\
  Z_\Pi &= \ket{\Pi}\bra{\Pi} - \ket{\Pi^\perp}\bra{\Pi^\perp}
\end{aligned}
\label{eq:grover_reflections}
\end{equation}
which reflect the state around the projection and $\psi$ axes, resulting in a rotation.
Implementing this rotation (the Grover operator) is the basis of Grover's algorithm for search~\cite{Grover2001}, and applying quantum phase estimation to the Grover operator was the first implementation of an amplitude estimation algorithm~\cite{Brassard2002}.
In 2019 multiple Ref.s~\cite{Aaronson2019, Suzuki2020, Wie2019} introduced methods of removing the Quantum Fourier Transform from the amplitude estimation algorithm, keeping the same query complexity.
Ref.~\cite{Aaronson2019} gave a proof for the lower bound query complexity, in~\cite{Suzuki2020} multiple samples of the Grover operator are taken, and Maximum Likelihood Estimation (MLE) is used to estimate the amplitude which improves the constant factor in query complexity, but has no proof.
In Ref.~\cite{Wie2019} the QFT of the original amplitude estimation algorithm is replaced with Hadamard tests, in a similar manner to the replacement of QFT with Hadamard tests in the Iterative Quantum Phase Estimation algorithm~\cite{Dobsicek2007a}.
Ref.~\cite{Grinko2021a} is another iterative amplitude estimation algorithm that improves on the constant factors of~\cite{Wie2019}, and has a rigorous proof.
Ref.~\cite{Callison2023} improves on MLE amplitude estimation~\cite{Suzuki2020} by showing a way to avoid values of the amplitude where MLE fails, and gives a method of achieving amplitude estimation with a given circuit depth, which is useful in early fault tolerant quantum computers, where the number of qubits and noise level restricts the overall circuit depth.


\subsection{Amplitude Estimation from Quantum Signal Processing}
\label{sub:amplitude_estimation_from_quantum_signal_processing} 

The scheme of~\cite{Rall2023} which we will use to implement amplitude estimation on a QLES correction vector requires access to arbitrary rotations around the reflection axes, $e^{i \phi_0 Z_\psi}, e^{i \phi_1 Z_\Pi}$.
Choice of phase factors, $\phi$, allows us to implement a family of polynomials on the amplitude, $P(a)$, which we can sample from with multiple shots of the circuit to get an estimator $|P(a)|^2$.
This sets up a quantum signal processing problem, where we are sampling from the polynomial $P$, where we measure in the $\{ \ket{\psi}, \ket{\psi^\perp} \}, (\{ \ket{\Pi}, \ket{\Pi^\perp} \})$ basis and return the state $\ket{\psi}(\ket{\Pi}) $ with probability $|P(a)|^2$, allowing us to sample from an estimator of $a$.
Each coin toss requires a number of oracle queries that depends on the degree of the polynomial, $O(\textrm{deg}(P) )$.
In Ref.~\cite{Rall2023} an algorithm that samples from Chebyshev polynomials is introduced, \textsc{ChebAE}, which has the lowest query complexity, due to the fact that Chebyshev polynomials have the greatest variation over the range $a \in [0, 1]$.

We have chosen a QSP algorithm as the basis of our amplitude estimation routine, yet as we have also used a QSVT implementation of the matrix inversion algorithm we must show that we are able to compose the two circuits together without incurring significant errors.
Thankfully there has been recent work~\cite{Mizuta2023a, Rossi2023a, Rossi2023b} we can use to show that the composition of the two functions for matrix inversion and amplitude amplification can be achieved when both are expressed as QSVT algorithms.
We will use the definitions from~\cite{Rossi2023a} of an \textit{embeddable} QSVT algorithm.

We require the matrix inversion subroutine to be an \textit{embeddable} QSVT algorithm.
An \textit{embeddable} algorithm allows us to nest the first algorithm as the signal operator of the outer algorithm, in general this is not possible for reasons we will briefly explain here, for a fuller review see Ref.~\cite{Rossi2023a}.
A QSVT algorithm is the lifted version of a QSP algorithm, where we act upon some 2 dimensional subspace of a multi-qubit Hilbert space, and we act upon the singular values of block-encoded matrices, instead of acting on the Hilbert space of a single qubit applying functions to a scalar value in QSP.
The 2 dimensional subspace of the QSVT  algorithm is then described by the projectors $\Pi_{\textrm{inv} }, \tilde{\Pi}_{\textrm{inv} }$\footnote{These are distinct from the projectors used in Equation~\ref{eq:projectors}, indicated by our use of the inv subscript to indicate matrix inversion.}, and the QSVT algorithm can be described as the circuit implementing,
\begin{equation}
  U_{\Phi_0} = e^{i \phi_1^0 (2 \tilde{\Pi}_{\textrm{inv}} - I )} U_0\prod_{j=1}^{(n -1) / 2} \left[ e^{i \phi_{2j - 1} (2 \Pi_{\textrm{inv}} - I)} U_0^\dagger e^{i \phi_{2j} (2 \tilde{\Pi}_{\textrm{inv}} - I )} U_0 \right]  
  \label{eq:qsvt_product}
\end{equation}
for even $n$ the number of phase factors in the inversion protocol, $\phi^0_j$.
Where $U_0$ implements the block encoding of the matrix to be inverted.

In the definition of nested QSVT protocols Ref.~\cite{Rossi2023a} makes the distinction between flatly nested and deeply nested QSVT protocols, where a flatly nested protocol shares its projectors with the outer QSVT routine. 
As we are nesting a matrix inversion algorithm within an amplitude estimation algorithm, we necessarily have different projectors so will focus upon deeply nested QSVT protocols.

To nest a QSVT protocol we replace the projectors of that protocol with a transformed projector, defined by the unitary transformation of the outer QSVT protocol, $U_{\Phi_1}$:
\begin{equation}
  \tilde{\Pi}  \rightarrow U_{\Phi_1} \tilde{\Pi}  U^\dagger_{\Phi_1}.
  \label{eq:transformed_qsvt_projectors}
\end{equation}

It is possible to choose any projector, $\Pi$, for the amplitude estimation algorithm, but the use of QSVT in the matrix inversion step indicates that we should project onto a tensor product of the all-$\ket{0}$ state of the flag qubits (which ensures we are in the top-left block of the block encoding), and some basis state corresponding to a location of a chosen correction peak.

The phase factors of the outer protocol are also restricted to be antisymmetric.
A list of phase factors is antisymmetric when it remains the same under reversal and negation, i.e. for a list of phase factors of length $2d$ the phase factors  are $\mathbf{\Phi}_1 = \left\{ \phi_0, \phi_1, \phi_2 \dots, \phi_d, - \phi_d, \dots, -\phi_2, - \phi_1, -\phi_0 \right\}$, and for an odd polynomial the central phase factor must be 0: $\mathbf{\Phi}_1 = \left\{ \phi_0, \phi_1, \phi_2 \dots, \phi_d, 0, - \phi_d, \dots, -\phi_2, - \phi_1, -\phi_0 \right\}$.
This restriction on the phase factors requires that the outer polynomial is real, and the antisymmetric list is unique for that polynomial.
The polynomial is therefore a rotation on the $XY$ plane of the Bloch sphere for all $x$, and anticommutes with the $Z$ rotations of the outer protocol.
If the polynomial were not real there would be a rotation that commutes with the $Z$ rotations of the inner protocol, and the composition of the function then loses information about this part of the inner function.
This argument is a brief explanation of the one in~\cite[Corollary II.4.1]{Rossi2023a}, which can be referenced for a full explanation.
In the follow-up work, Ref.~\cite{Rossi2023b} the authors discuss extensions to a greater family of protocols.

For our purposes all that remains is to show that the outer protocol, that discussed in~\cite{Rall2023}, implements a real polynomial and therefore that an antisymmetric list of phase factors exists.
Thankfully, the algorithm we will be using from Ref.~\cite{Rall2023} is \textsc{ChebAE} which implements amplitude estimation using Chebyshev polynomials as the $P(a)$ transformation, and as the Chebyshev polynomials are real we can guarantee that an antisymmetric list of phase factors exist, and therefore that we can embed the matrix inversion QSVT protocol into the \textsc{ChebAE} protocol.

\subsection{Resource Estimation of the \textsc{ChebAE} algorithm}
\label{sub:resource_estimation_of_the_chebae_algorithm} 

In Algorithms~\ref{alg:cheb_ae} and~\ref{alg:find_next_cheb} we have re-produced the \textsc{ChebAE} and \textsc{FindNextCheb} algorithms from Ref.~\cite{Rall2023} so that this paper is complete.
Ref.~\cite{Rall2023} provides an average query complexity to the oracles that implement rotations in the \textsc{ChebAE} algorithm, but for a full resource estimation we need to determine the cost of a single oracle in our matrix inversion setting.
Instead of the reflection operators, $Z_\psi, Z_\Pi$ in Equation~\ref{eq:grover_reflections} we require access to the rotations about the axis defined by these reflections, $e^{i \phi_j Z_\psi}, e^{i \phi_j \Pi}$.
Following the exposition in \cite{Rall2023}, we can implement $Z_{0^n} = 2 \ket{0^n} \bra{0^n} - I$ with Toffoli gates and a $Z$ gate, and therefore $e^{i \phi_j Z_{0^n}}$ with the same number of Toffoli gates and a $Z$ rotation.
Implementation of $e^{i \phi_j Z_\psi}$ is then done via $U_\psi e^{i \phi_j Z_{0^n}} U^\dagger_\psi$.
For the rotation $e^{i \phi_j Z_\Pi}$ we make a similar argument, that instead of trying to implement $Z_{0^n}$, $Z_\Pi$ is a projection onto some other, known, basis state.
This projector has an exact cost that differs for the exact basis state considered, but is generally $\mathcal{O}(n)$ Toffoli gates.
The rotation can be implemented by replacing the final $Z$ gate with a $R_z$ gate.
Therefore  $e^{i \phi_j Z_\Pi}$ costs $\mathcal{O}(n)$ Toffoli gates and one rotation gate, and $e^{i \phi_j Z_\psi}$ is $2 \times$ the cost of the matrix inversion sub-routine, $\mathcal{O}(n)$ Toffoli gates and a single rotation gate. 

\begin{algorithm}[H]
  \floatname{algorithm}{Algorithm}
  \algrenewcommand\algorithmicrequire{\textbf{Input: }}
  \algrenewcommand\algorithmicensure{\textbf{Output: }}
  \caption{\textsc{ChebAE}}\label{alg:cheb_ae}
  \begin{algorithmic}[1]
    \Require $\epsilon_{\textrm{meas} }$: Required accuracy, $\ket{\psi}$: State to be measured, $\Pi$: Projector to measure onto
    \Require Hyperparameters: $r = 2, \nu = 8, N_{\textrm{shots} } = 100$ \Comment{Hyperparameters, defaults found in~\cite{Rall2023}}
    \Ensure $\hat{a}$: Estimation of the amplitude $a$.
    \State $T \gets \lceil \log_r( (2\epsilon_{\textrm{meas}} )^{-1} ) \rceil$ \Comment{$T$ is the upper bound on the number of confidence intervals needed.}
    \State $\epsilon_{\textrm{max}}^{p} \gets \textsc{ClopperPearson}(N_{\text{shots}}, 1 - \frac{\delta}{T})$ \Comment{Use the Clopper-Pearson method to estimate the largest possible error on the estimate of the bias of a coin with $N_\text{shots}$ flips and confidence $1-\delta/T$.}
    \State $[a_\text{min},a_\text{max}] \gets [0,1]$
    \State $n_\text{heads},n_\text{flips} \gets 0,0$
    \State $d \gets 1$
    
    \While{$a_\text{max} - a_\text{min} \geq 2\epsilon_{\textrm{meas} }$}
      \State $d_{\text{new}} \gets \textsc{FindNextCheb}(a_\text{min},a_\text{max})$ 
      \If{$d_\text{new} \geq rd$}
        \State $n_\text{heads},n_\text{flips} \gets 0,0$
        \State $d \gets d_\text{new}$
      \EndIf
      \If{$\epsilon^{p}_\text{max} \cdot \frac{a_\text{max} - a_\text{min}}{|T_d(a_\text{max}) - T_d(a_\text{min})|} \leq \epsilon_{\textrm{meas} }\nu$}
        \State $N_{\textrm{samples}} = 1$ \Comment{``Late stages"}
      \Else
        \State $N_{\textrm{samples}} = N_{\textrm{shots} }$ \Comment{``Early stages"}
      \EndIf
      \For{$i \gets 0, i \leq N_{\textrm{samples}}, i \gets i + 1$}
      \State $T_d(a) = \textrm{cos}(d \textrm{ arccos}(a) ) \gets \textsc{QSP}(\psi, \Pi)$\Comment{Use the Quantum Signal Processing algorithm defined in~\cite{Rall2023} to apply the Chebyshev polynomial $T_d$ to $\Pi \ket{\psi}$ so we can sample from $T_d(a)$ in the next step.}
        \State $s \gets \textsc{Sample}(T_d(a))$
        \State $n_{\textrm{flips}} \gets n_{\textrm{flips}} + 1$
        \If{$s = 1$}
         \State $n_{\textrm{heads}} \gets n_{\textrm{heads}} + 1$
        \EndIf
      \EndFor
    \State $[p_\text{min},p_\text{max}] \gets \textsc{ClopperPearson}(n_{\textrm{flips}}, \frac{\delta}{T})$ \Comment{Compute a $\delta/T$ confidence interval  on $|T_d(a)|^2$ using the Clopper-Pearson method \cite{Clopper1934}.}
  \State $[p_\text{min},p_\text{max}] \gets [a^*_\text{min},a^*_\text{max}]$ \Comment{Find new bounds, $a^*$, by choosing the values of $a$ that correspond to the sampled $p_{\textrm{min}},p_{\textrm{max}}$. This is shown in Figure 5 of~\cite{Rall2023}.}
        \State $[a_\text{min},a_\text{max}] \gets [a^*_\text{min},a^*_\text{max}]\cap [a_\text{min},a_\text{max}]$
    \EndWhile
    \State $\hat a \gets$ midpoint of $[a_\text{min},a_\text{max}]$
    \State \textbf{return} $\hat{a}$
  \end{algorithmic}
\end{algorithm}

The \textsc{FindNextCheb} subroutine finds the highest degree Chebyshev polynomial $|T_d(a)|^2$ that is invertible in the $[a_\text{min},a_\text{max}]$ region.

\begin{algorithm}[H]
  \floatname{algorithm}{Algorithm}
  \algrenewcommand\algorithmicrequire{\textbf{Input: }}
  \algrenewcommand\algorithmicensure{\textbf{Output: }}
  \caption{\textsc{FindNextCheb}}\label{alg:find_next_cheb}
  \begin{algorithmic}[1]
    \Require $a_{i\text{max}}, a_{i\text{min}}$\Comment{Current estimation for the upper and lower bound of $a$}
    \Ensure $d$: New degree of Chebyshev polynomial.
        \State $[\theta_\text{min},\theta_\text{max}] \gets [\arccos(a_\text{max}),\arccos(a_\text{min})]$
    \State $d \gets \left\lfloor\frac{\pi}{2} (\theta_\text{max}-\theta_\text{min})^{-1} \right\rfloor$.
  \While{$\cos^2(d \theta)$ has no extrema for $\theta \in [\theta_\text{min},\theta_\text{max}]$}
   \State $d \gets d - 1$ \Comment{This is equivalent to $\left\lfloor \frac{2 }{\pi}d\theta_\text{min} \right\rfloor = \left\lfloor \frac{2 }{\pi}d\theta_\text{max} \right\rfloor$.}
  \EndWhile
    \State \textbf{return} $d$
  \end{algorithmic}
\end{algorithm}

\subsubsection{A Note on Accuracy, $\epsilon$}
\label{sec:a_note_on_accuracy_epsilon}

In what follows the term accuracy will have some different meanings, which we briefly discuss here to avoid confusion.
We are using essentially three subroutines to solve the CFD problem, the classical CFD loop, the QSVT matrix inversion algorithm, and the amplitude estimation algorithm.
Each of these defines its own accuracy component.
We will denote the accuracy, or tolerance, of the CFD algorithm as $\epsilon_{\textrm{tol} }$, where when $\textrm{max}(|a_i|) \leq \epsilon_{\textrm{tol} }$ for all $a_i$ in the corrections vector we stop the algorithm and return the result.
The accuracy of the amplitude estimation we denote $\epsilon_{\textrm{meas}}$ which is the accuracy of measuring a single amplitude.
Formally, we say that an amplitude estimation algorithm samples from a random variable $\hat{a}$ satisfying,
\begin{equation}
  Pr [|\hat{a} - |a| | \geq \epsilon_{\textrm{meas} }] \leq \delta,
  \label{eq:epsilon_meas}
\end{equation}
for some probability of failure, $\delta$.
Finally, there is an accuracy component to the matrix inversion, $\epsilon_{\textrm{inv} }$, which we use to combine the accuracy of the polynomial approximating $f(A) = A^{-1}$, and the accuracy of rotation gates in a fault tolerant quantum computer.


\subsubsection{Empirical Query Complexity}
\label{sec:empirical_query_complexity} 

We use Empirical Claim 18 of the \textsc{ChebAE} algorithm from~\cite{Rall2023}, the \textsc{ChebAE} samples from a random variable $\hat{a}$ satisfying Eqn.~\ref{eq:epsilon_meas}.
For each iteration of the amplitude algorithm, a number of samples need to be taken to get a good input to the next iteration of the algorithm.

The number of samples at each iteration, and the number of overall iterations can be combined to give an overall query complexity of $Q_\Pi$.
$Q_\Pi$ is the number of times we implement the projection operator in the circuit, and each time the projector is queried we need to implement the unitary $U$ once and the unitary  $U^\dagger$ once.
For the \textsc{ChebAE} algorithm $Q_\Pi$ has been modelled empirically:
\begin{equation}
  \langle Q_\Pi \rangle \approx \displaystyle\frac{A}{\epsilon_{\textrm{meas} }} \textrm{ln} \left( B \textrm{ln}\left( \frac{1}{\epsilon_{\textrm{meas} }} \right)   \right)
  \label{eq:query_complexity}
\end{equation}
where $A = 1.71, B = 2.18$ 
for some reasonable parameters, $a = 0.5, \delta = 0.05, \epsilon_{\textrm{meas}} \in [10^{-3}, 10^{-6}]$ \footnote{Simulated examples show that the query complexity for 95\% of trials deviate from $\langle Q_\Pi \rangle$ by a maximum of $3.15\%$. For other values of $a$ this deviation can be larger. As we use the value for $a=0.5$ in the rest of this paper the maximum deviation of $3.15\%$ is reasonable.}.
We note that the number of query calls is dependent on $\delta, \epsilon_{\textrm{meas}},$ and $a$. Whereas the choice of $\delta, \epsilon_{\textrm{meas}}$ is free the absolute value of the amplitude is not known \textit{a priori}.
We have re-calculated the query complexity for different values of $a$, keeping $\delta = 0.05, \epsilon_{\textrm{meas}} \in [10^{-3}, 10^{-6}]$, and focus on the value of $A$ which is the dominant factor.
The maximum value of $A$ occurs at small $a=0.01$, which will be unimportant for the algorithm proposed here, as we will only record the largest peaks.
The highest value of $A$ for peaks which will be relevant here is that reported by~\cite{Rall2023}, $A=1.71$ at $a=0.5$, so for the resource estimation here we will use the $\langle Q_\Pi \rangle$ equation from~\cite{Rall2023}, with $A=1.71, B=2.18$.

\subsection{Modification for the sign of $a$} 
\label{sub:modification_for_the_sign_of_a}

The amplitude estimation algorithm above allows us to sample from the variable $\hat{a} $ with probability as in Equation~\ref{eq:epsilon_meas}, $Pr = \left[ | \hat{a} - |a| \geq \epsilon_{\textrm{meas} } |  \right]$ \cite[Theorem 6]{Rall2023}, but this does not give the sign of the correction, which we can see from Figure~\ref{fig:output_vec} is required.
Therefore, we use the idea from~\cite{Manzano2023} of applying an iteration, $i$, dependent shift $\pm b_i$ to $a$ that gives us information on the sign of $a$.
We must then initialise separate \textsc{ChebAE} algorithms for the positive and negative shifts, doubling the query complexity.
This algorithm is described in Algorithm~\ref{alg:final_algorithm}.

We modify the original algorithm by taking a measurement of
\begin{equation}
  \ket{\psi_i}_\pm = (a \pm b_i) \ket{\Pi} + \sqrt{1 - (a \pm b_i)^2} \ket{\Pi^\perp},
  \label{eq:psi_pm_projector}
\end{equation}
As $a, b_i \in \mathbb{R}$ we can show that
\begin{equation}
    a = \frac{(a + b_i)^2 - (a - b_i)^2}{4 b_i}.
\end{equation}
In every iteration of real amplitude estimation we build $\hat{a}_i$, an estimation of $a$ using the probabilities of obtaining $\ket{\Pi}$ when measuring $\ket{\psi_+}, \ket{\psi_-}$, $\hat{p}_+, \hat{p}_-$:
\begin{equation}
    \hat{a}_i = \frac{\hat{p}_+ - \hat{p}_-}{4 b_i}
    \label{eq:hat_a_sign}
\end{equation}
Unlike the estimator $\hat{a}$ in Subsection~\ref{sec:a_note_on_accuracy_epsilon}, this estimator can take negative values, $\hat{a}_i \in (-1, 1)$.
This means that we can estimate the sign of $a$ as well as its magnitude.
The first value of the shift, $b_0$ can be arbitrary with a defined sign, and subsequent iteration set the shift to be negative the previous lower bound: $b_{i + 1} = -a_i^{\textrm{min} }$.
This ensures that bounds on the estimation of $a$ tighten with each iteration, and that $a$ has a defined sign.
More details on setting $b_i$ for subsequent iterations can be found in Ref.~\cite{Manzano2023}.

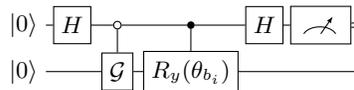
\begin{figure}[ht]
  \begin{center}
    \begin{tikzpicture}
      \begin{yquant}
        qubit {$\ket{0}$} a[1];
        qubit {$\ket{0}$} s[1];

        h a;
        box {$\mathcal{G}$} (s) ~ a[0];
        box {$R_y(\theta_{b_i})$}  s[0] |  a[0];
        h a;
        measure a;

      \end{yquant}
    \end{tikzpicture}
  \end{center}
  \caption{Circuit used to implement the shifted states $\ket{\psi_\pm}$ using a controlled implementation of $\mathcal{G}$. Here, $\cos{\theta_b} = b$}
  \label{fig:shifted_states_circuit}
\end{figure}

However, we still need to know how to implement the $\ket{\psi_\pm}$ operators.
We will use the symbol $\mathcal{G}$ to describe the overall effect of one iteration of the amplitude estimation algorithm, e.g. $\mathcal{G} = i^{2k + 1}e^{-i \phi_0 Z_\Pi} \prod_{j=1}^k (e^{i \phi_{2j -1}Z_\psi} e^{i \phi_{2j} Z_\Pi}) e^{i \phi_{2k + 1}Z_\psi}$\footnote{Note that the specific form of $\mathcal{G}$ depends upon the parity of the amplitude estimation polynomial and the input state, $\ket{\Psi}, \ket{\Pi}$.}.
The simplest method requires two implementations of the amplitude estimation algorithm $\mathcal{G} \ket{0} \approx  a \ket{0} + \sqrt{1 - a^2} \ket{0^\perp}$ for each shift value, by applying the shift onto the measured register at the end of the circuit.
However, we can use a Hadamard gate and controlled implementation of $\mathcal{G}$ to achieve the same result, as  in Figure~\ref{fig:shifted_states_circuit}.

The final state before measurement is then:
\begin{equation}
  \frac{a + \cos{\theta_{b_i}}}{2} \ket{0 0} + \frac{-a + \cos{\theta_{b_i}}}{2} \ket{1 0} + \dots
  \label{eq:shifted_aa_state}
\end{equation}
where the whole state in the system qubits has been omitted for brevity.
This lets us estimate the amplitude in $\ket{\psi_\pm}$, depending on the value of the ancilla qubit.

Application of the shift $b_i$ to the amplitude will affect the average query complexity $\langle Q_\Pi \rangle $, due to the dependence of $\langle Q_\Pi \rangle$ on $a$; in our resource estimation we choose the constants in Eqn.~\ref{eq:query_complexity} from $a = 0.5$, which results in a higher query complexity.
However, as each circuit now returns an estimator for either $a + b_i$ or $a - b_i$,  and has a separate \textsc{ChebAE} instance, the required accuracy of each can be halved.
To complete a resource estimation of amplitude estimation that also required us to estimate the sign, we must add the resources required to implement a controlled $\mathcal{G}$ operator, and a $cR_y$ operator.
As we are using QSP and QSVT algorithms here, to create a control $\mathcal{G}$ operator, we only need to use controlled-rotation gates to implement the phase factors of the algorithm, as without the phase shift applied by these gates the operators in the rest of the circuit cancel out, this can be seen from Ref.~\cite[Fig. 1]{Gilyen2019}.
The $cR_y$ operator requires that we project into the subspace $\left\{   \ket{0}, \ket{0^\perp} \right\}$, and apply the controlled $R_y$ gate on this subspace.
The cost of projecting into the subspace has been discussed above, it is the cost of $Z_\Pi$, except now we know that $\Pi = \ket{0}\bra{0}$, so the projection cost is $\mathcal{O}(n)$ Toffolis, and a controlled rotation is twice the T gate cost of an uncontrolled rotation.

\begin{algorithm}[H]
  \floatname{algorithm}{Algorithm}
  \algrenewcommand\algorithmicrequire{\textbf{Input: }}
  \algrenewcommand\algorithmicensure{\textbf{Output: }}
  \caption{Real Amplitude Estimation based on QSP}\label{alg:final_algorithm}
  \begin{algorithmic}[1]
  \Require $\phi_\psi$: Phase factors describing the matrix inversion algorithm, $\Pi$: projector on the subspace to be measured, $a = \Pi \ket{\psi}$, $\epsilon_{\textrm{meas}}$: desired accuracy, $\delta$: Probability of failure.

  \Ensure $\tilde{a}$: Sample from the estimator $\textrm{Pr} \left[ \left|\hat{a} - a \right|  \geq \epsilon_{\textrm{meas}} \right] \leq \delta$ 
  
  \State{$b_0$}\Comment{Choose an arbitrary shift with a defined sign}
  \State $i \gets 0$
  \State $\epsilon^a_{\textrm{meas}, i} \gets 2$
  \While{$\epsilon_{\textrm{meas}, i} \geq \epsilon_{\textrm{meas} } $}
  \State{$\mathcal{G},  \gets \textsc{QSP}(\Pi, \hat{a})$} \Comment{Determine circuit to implement $T(d)(a)$ from subroutine \textsc{QSP}}
  \State Initialize $\textsc{ChebAE}_+, \textsc{ChebAE}_-$ \Comment{Initialise two \textsc{ChebAE} subroutines for calculating the value of $a + b_i, a - b_i$}
  \State $N_{\textrm{samples} } \gets \textsc{ChebAE}(\Pi, \psi, \epsilon_{\textrm{meas} })$ \Comment{Get the number of samples to take from the \textsc{ChebAE} subroutine.}
  \For{$j = 0, j \leq 2 N_{\textrm{samples} }, j \gets j + 1$}
    \State{$m_j \gets \textsc{Measure}(\textrm{qubit}_0)$} \Comment{Measure the ancilla qubit in Figure~\ref{fig:shifted_states_circuit}}
    \If{$m_j = 0$}
      \State $\hat{p}_+ \gets  \textsc{ChebAE}_+(\Pi, \psi, \epsilon_{\textrm{meas} })$ \Comment{If the outcome of the ancilla qubit is 0, we are sampling from  $T_d(a + b_i)$ and should update $n_{\textrm{heads} }, n_{\textrm{flips} }$ in $\textsc{ChebAE}_+$ subroutine.}
    \Else
      \State $\hat{p}_+ \gets  \textsc{ChebAE}_-(\Pi, \psi, \epsilon_{\textrm{meas} })$ \Comment{Otherwise we sample from $T_d(a - b_i)$, and update $n_{\textrm{heads} }, n_{\textrm{flips} }$ in the $\textsc{ChebAE}_-$ subroutine.}
  \EndIf
  \EndFor
  \State{$\hat{a}_i = \frac{\hat{p}_+ - \hat{p}}{4b_i}$}\Comment{The estimation of $a$ is given by the result of the circuit in Figure~\ref{fig:shifted_states_circuit} }
  \State{$a_i^{\textrm{max}} \gets \textrm{min}\left( \frac{\hat{p}_+ - \hat{p}_-}{4 b_i} + \frac{\epsilon^p_{\textrm{meas}, 1}}{2 b_i}, 1\right)  $}
  \State{$a_i^{\textrm{min}} \gets \textrm{max}\left( \frac{\hat{p}_+ - \hat{p}_-}{4 b_i} - \frac{\epsilon^p_{\textrm{meas}, 1}}{2 b_i}, -1\right)  $}
  \State{$a_i = \frac{a_i^{\textrm{max}} + a_i^{\textrm{min} } }{2}$}
\State{$\epsilon^a_{\textrm{meas}, i} = \frac{a_i^{\textrm{max}} - a_i^{\textrm{min} } }{2}$}
  \State $b_{i + 1} = -a_{\textrm{min}} $
  \EndWhile
  \State \textbf{return} $\hat{a}$
  \end{algorithmic}
\end{algorithm}


\subsection{Overall Cost of Amplitude Amplification} 
\label{sub:overall_cost_of_amplitude_amplification}

In the remainder of this paper we will focus upon an amplitude estimation algorithm that is the combination of amplitude estimation from QSP~\cite{Rall2023} with the Real Quantum Amplitude Estimation (RQAE) algorithm of~\cite{Manzano2023}.
The Quantum Signal Processing algorithm is used to transform $a = \Pi \ket{\psi}$ into $T_d(a)$, a Chebyshev polynomial of the amplitude $a$, and in Algorithm~\ref{alg:cheb_ae} we detail how the \textsc{ChebAE} algorithm is used to estimate $a$ given this transformation.
As we require the sign of the amplitude, we follow RQAE and introduce a shift, $b_i$ to the state.
We must therefore follow the \textsc{ChebAE} algorithm for two amplitudes, sampling from $T_d(a + b_i)$, and $T_d(a - b_i)$.
This procedure is detailed in Algorithm~\ref{alg:final_algorithm}.
We can now use the RQAE algorithm~\cite{Manzano2023} to introduce a shift to the amplitude.
We initialise two instances of the \textsc{ChebAE} algorithm to estimate $(a + b_i), (a - b_i)$ and update the shift according to RQAE.

By inspecting the structure of the new algorithm, we can use the empirical query complexity given by~\cite{Rall2023} and the circuit structure from~\cite{Manzano2023} to estimate the total resources required for the new algorithm.
We also require the number of qubits in that circuit, $n$.
Whilst we aim for a concrete resource estimation we do not know the exact value of $\Pi$, so therefore must use some asymptotic estimates, e.g. $\mathcal{O}(n)$.
We will approximate these values as $n$.
Using Equation~\ref{eq:query_complexity}, for amplitude estimation with QSP, Algorithm~\ref{alg:final_algorithm} requires two instances of \textsc{ChebAE} for each shift value, meaning we must double the query complexity of Equation~\ref{eq:query_complexity}.
\begin{equation}
  \langle Q_\Pi \rangle \approx_{3.15 \%} 2 \times \displaystyle\frac{1.71}{\epsilon_{\textrm{meas} }} \textrm{ln} \left( 2.08 \textrm{ln}\left( \frac{1}{\epsilon_{\textrm{meas} }} \right)   \right) 
  \label{eq:query_complexity_2}
\end{equation}

We can now estimate the cost of one of these queries based upon the cost of $U_\psi = \ket{\psi}$.
This is a QSVT algorithm with a known degree, $d$, and known non-Clifford cost of the  block encoding, $\mathcal{M}_{C}$.
We know that one iteration of amplitude estimation requires a call to $U_\psi$, and one call to $U_\psi^\dagger$ so the degree of the new QSVT algorithm is twice that of the inversion algorithm, which doubles the non-Clifford cost.
The implementation on $Z_\Pi$ requires $\mathcal{O}(n)$ Toffoli gates, we do not know exactly as the choice of projector is not set.
Finally, to estimate the sign of the amplitude as per~\cite{Manzano2023} we need to implement the $\bar{c}\mathcal{G}, cR_y$ operators.
The $cR_y$ gate costs $\mathcal{O}(n)$ Toffoli gates, and the T cost of a controlled rotation, which is double the cost of a single rotation gate~\cite{Shende2006a}.
The cost of $\bar{c}-\mathcal{G}$ is the cost of one iteration of the amplitude estimation algorithm, with the T cost of the phase factor gates doubled to account for the control.

The non-Clifford gate cost of a single query ($Q_\Pi$) of the amplitude estimation algorithm is given by:
\begin{equation}
  \textsc{Cost} \left(Q_\Pi \right) = 2d \mathcal{M}_{\textrm{C} } + 2n \textrm{Toff} + (4d  +2)\mathcal{R}_T,
  \label{eq:query_cost}
\end{equation}
where $d$ is the degree of the matrix inversion polynomial, $\mathcal{M}_\textrm{C}$ is the non-Clifford cost of the matrix inversion block encoding and projection operators, and  $\mathcal{R}_T$ is the T- cost of a single rotation.

The space cost of the full routine is
\begin{equation}
  \textsc{Space} \left(Q_\Pi \right)\ = \mathcal{M}_{\textrm{qb} } + 2,
  \label{eq:space_cost}
\end{equation}
where $\mathcal{M}_{\textrm{qb} }$ is the qubit cost of the matrix inversion sub-routine.
Equations~\ref{eq:query_cost} and~\ref{eq:space_cost} are combined for to give a full resource estimation for the CFD use-case in Table~\ref{tab:amplitude_encoding_resource_requrements} and the Toeplitz matrix in Table~\ref{tab:toeplitz_resource_requrements}.

In the remainder of this paper we will focus upon the CFD use-case of matrix inversion via QSVT.


\section{CFD Method} 
\label{sec:cfd_method}

In this work Algorithm~\ref{alg:final_algorithm} for estimating the sign and value of an amplitude, based on QSP amplitude estimation is applicable to any state generated via a QSVT algorithm.
Algorithm~\ref{alg:measurement_scheme} is applicable only to the situation where a QSVT algorithm is used to generate the corrections vectors within a Computational Fluid Dynamics (CFD) loop.
In this section we will briefly outline the calculations simulated in Section~\ref{sec:results}.

Computational Fluid Dynamics (CFD) is used to simulate and engineer many modern components that interact with fluids, e.g. the flow of gas through a turbine engine~\cite{Hills2007}.
The Navier-Stokes equations describe the conservation of mass, momentum, and energy within the flow of a fluid.
The equations are highly non-linear, however, many CFD solvers adopt an iterative approach whereby the equations are linearised and then solved to provide an update to the non-linear flow field. Each linearised system can be expressed in the form of residual equations:
\begin{equation}
\begin{aligned}
  A^{n} \mathbf{\delta x^{n+1}} &= \mathbf{\delta b^{n}} = 
  \mathbf{b^{n}} - A^{n} \mathbf{x^{n}} \\
  \mathbf{x^{n+1}} &= \mathbf{x^{n}} + \mathbf{\delta x^{n+1}}
  \label{eq:matrix_inversion_nonl}
\end{aligned}
\end{equation}
where $A^n$ and $\mathbf{b^n}$ depend on  $\mathbf{x^n}$ and
the iterations begin with a initial guess for the flow field at $n=0$.
The iterations are repeated until the $L_2$ norm of the non-linear equation errors, $||\mathbf{\delta b^{n}}||$, falls below a specified threshold, $\epsilon_{\textrm{tol} }$. 
In the work here, we will use $\epsilon_{\textrm{tol}} = 1 \times 10^{-9}$.
Note that to use a QLES the state vectors $\mathbf{\delta x}$ and $\mathbf{\delta b}$ must be normalised. This is why the vectors in Figures~\ref{fig:output_vec_a} and ~\ref{fig:output_vec_b} do not show the reduction in the $L_2$ residual error.

In this work, we use two different solvers, either the Semi-Implicit Method for Pressure Linked Equations (SIMPLE) solver, a derivation of which can be found in the Appendix of Ref.~\cite{Lapworth2022b}, or a coupled solver.
In the former, the linearised momentum and mass conservation equations are solved separately, leading to a Poisson type equation for corrections to the pressure field. 
In the latter, the linearised equations are combined in a single, coupled, matrix system \cite{Lapworth2022c}.
This has the benefit of providing faster convergence.
In this work we solve the CFD problem for a 1D convergent-divergent channel, sometimes referred to as a  nozzle, and the discretization scheme used is also described in~\cite{Lapworth2022b}.
We can consider two types of fluid, incompressible and compressible: for the former the density is constant, whereas, for the latter the Perfect Gas Law and energy conservation are used to relate pressure, density and temperature.


\section{Measurement Scheme for Corrections Vectors}
\label{sec:measurement_scheme_for_corrections_vectors} 

We have now covered all of the sub-routines that constitute the measurement scheme proposed here.
In this section we will restrict ourselves to the CFD use case, and the structure of corrections vectors returned in this case.
Figure~\ref{fig:output_vec} shows two example corrections vectors from a CFD loop, in the early and late stages of the algorithm, as there are only a small number of peaks, we propose an algorithm that measures only basis vectors that have an absolute amplitude grater than some cutoff proportion, $\alpha$ of the maximum amplitude: $|a| \geq \alpha |a_{\textrm{max} }|$.
In Section~\ref{sub:effect_of_noisy_measurements_on_the_qles_solver} we  will simulate this scheme to study the effect of this cutoff on the total CFD loop.

The first part of the algorithm is using QSVT to invert the input matrix and apply it to the input vector, creating the state $\ket{\psi}$. 
The output vector will be stored in wavefunction represented by the top-left block of the block encoding.
We then choose a  projection, $\Pi$, to measure which must include a projection of the QSVT flag qubits into the $\ket{0}^{\otimes \textrm{flag} }$ state, which ensures we are in the top-left block.
We can then use Algorithm~\ref{alg:final_algorithm} to estimate the sign and value of $a = \Pi \ket{\psi}$.

Without prior knowledge of the setting, we may be forced to measure all amplitudes in the output vector using this method, however, we see from Figure~\ref{fig:output_vec} that there are a smaller number of dominant peaks, spread out over a number of amplitudes.
We introduce $\alpha$, the cutoff, which is the absolute value of an amplitude, relative to the (absolute) height of the maximum amplitude.
We will not measure any basis vectors in the final wavefunction which are below the cutoff.
For a correction vector $\mathbf{a}$, where $a_{\textrm{max}} = \textrm{max} |\mathbf{a}|$, for all elements in the vector, $a_i, i \in {0, 1, \dots 2^n}$:
\begin{equation}
  a_i  =
  \begin{cases}
    a_i & \textrm{ if } |a_i| > \alpha \; | a_{\textrm{max}} | \\
    0 & \textrm{ otherwise.}  
  \end{cases}
  \label{eq:alpha_cutoff}
\end{equation}
Consider Figure~\ref{fig:output_vec_a}, where the maximum peak has a height of $\approx 0.5$, and there are two smaller, negative peaks with a maximum height of $\approx 0.3$, choosing a cutoff of $\alpha = 0.5$ would require the measurement of all three peaks, whereas with a cutoff $\alpha = 0.9$, only the most dominant peak is measured.

Whilst this is a strategy for reducing the number of $a_i$ that must be measured, we still do not know \textit{a priori} which basis vectors to choose.
We must therefore introduce the `burn-in' period, where we are required to measure the output vector without any amplitude estimation for a number of times to determine which basis vectors record the highest peaks.
The number of measurements in the burn-in period is then determined by the accuracy we desire, i.e. we wish to distinguish $|a_i| \approx 0$ from $|a_i| \approx \alpha a_{\textrm{max} }$, which implies we need to measure $\frac{1}{(\alpha a_{\textrm{max} })^2}$ times.
This does not give us an \textit{a priori} number of measurements either, as $a_{\textrm{max} } \in [\frac{1}{2^n}, 1]$\footnote{Lower bound is all basis vectors are measured with equal probability, and upper bound is only one amplitude present in the final wavefunction.}.
The absolute number of measurements required is low compared to the overall runtime of the algorithm, yet below we present a method for estimating the required number of measurements.

\subsection{Modelling $a_{\textrm{max}}$}
\label{sub:modelling_a_max} 

To tackle this, we propose an algorithm specifically for the `burn-in' period, designed to determine if $a_{\textrm{max} }$ is low or high.
Begin by taking samples of the final measurement, called $N_S$, and record $N_{\textrm{new} }$ the number of unique measurement outcomes seen, the ratio of these numbers $r = \frac{N_{\textrm{new}}}{N_S}$ gives an indication of the value of $a_{\textrm{max}}$.
A rough model of the behaviour can be thought of as a wavefunction which is uniform everywhere except a small number of large peaks, which have outsized amplitudes compared to the rest of the state.
Then, if the value of $a_{\textrm{max} }$ is very high compared to the background, we will measure the basis vector(s) corresponding to $a_{\textrm{max}}$ and other high peaks relatively more times than the background basis vectors, and $r$ will be small.
If $a_{\textrm{max} }$ is closer to the background values, $r$ will be higher.

We have taken corrections vectors from classical CFD calculations and separated the recorded $|a_{\textrm{max} }|$ into decile bins.
We have averaged over all the vectors recorded for both solvers, compressible and incompressible fluids for 3 - 8 qubit systems and calculated a linear relationship between $r$ and $a_{\textrm{max} }$, which is:
\begin{equation}
  r = -0.86 a_{\textrm{max} }  + 0.80.
  \label{eq:a_max_r}
\end{equation}
So for the `burn-in' period with a number of samples $N_S \geq 2$, given the wavefunction behaves roughly as the ones modelled here (which is the case for CFD corrections) we can find a value for $r$ and use this model to estimate $a_{\textrm{max} }$ and therefore the total number of shots to take in the `burn-in' period, $\frac{1}{(\alpha |a_{\textrm{max} }|)^2}$
Additional details on the modelling of $a_{\textrm{max} }$ are given in Appendix~\ref{app:modelling_a_max}


\subsection{Cutoff Algorithm}
\label{sub:Cutoff Algorithm} 

The novel measurement scheme is detailed in Algorithm~\ref{alg:measurement_scheme}.

\begin{algorithm}[H]
  \floatname{algorithm}{Algorithm}
  \algrenewcommand\algorithmicrequire{\textbf{Input: }}
  \algrenewcommand\algorithmicensure{\textbf{Output: }}
  \caption{Cutoff Algorithm}\label{alg:measurement_scheme}
  \begin{algorithmic}[1]
    \Require $U_\psi$: Block encoding of inversion matrix, $N_{\textrm{max} }$: Maximum number of shots in the burn-in period, $\alpha$: Cutoff value
    \Ensure $\left\{ a_{\Pi_j} \right\}, \left\{ \Pi_j \right\}$: Signed amplitude and basis vectors of most significant corrections.
    \State $\Pi \gets \ket{0}^{\otimes \textrm{flag} }$
    \For{$i \gets 0, i \leq N_{\textrm{max}}$, i += 1}
    \State $\ket{\psi_{AE}} \gets \textsc{AmplitudeAmplification}(\ket{\psi}, \Pi)$ \Comment{Use amplitude amplification to ensure we are in the top-left block}
    \State $\Pi_i \gets \textsc{Measure}(\ket{\psi_{AE}})$ \Comment{Measure the state and record the output}
    \State $N_S \gets N_S + 1$
    \If{$\Pi_i \notin \left\{ \Pi_j \right\} $}
      \State $N_{\textrm{new} } \gets N_{\textrm{new}} + 1$
      \State $\left\{ \Pi_j \right\} + \Pi_i$ \Comment{Add the unseen measurement to the measurement record}
      \State $\left\{ a_j \right\} + a_i \gets 1$ \Comment{Add one measurement to the estimation of $a_i$}
    \Else
      \State
      \State $\left\{ a_j \right\} + a_i \gets 1 $ \Comment{Add one measurement to the estimation of $a_i$}
    \EndIf
    \State $r \gets \frac{N_{\textrm{new} }}{N_S}$
    \If{$i \% 10 = 0$}
      \State $a_{\textrm{max}} \gets \textsc{Solve}\left( \text{Eqn.\ref{eq:a_max_r}} \right)$ \Comment{Solve Equation~\ref{eq:a_max_r} to get an estimate of $a_{\textrm{max}}$}
      \If{$i \geq \frac{1}{(\alpha a_{\textrm{max}})^2}$}
        \State $\left\{ a_j \right\}  / i$ \Comment{Normalise estimated amplitudes}
        \State \textbf{break}
      \EndIf
    \EndIf
    \State $\left\{ a_j \right\}  / i$ \Comment{Normalise estimated amplitudes}
    \EndFor
    \State $a_{\textrm{max}} \gets \textrm{max}\left\{ |a_i| \right\} $
    \For{$\Pi_i \in \left\{ \Pi_j \right\}$}
    \If{$|a_i| \leq \alpha |a_{\textrm{max}}|$} \Comment{Skip any basis vectors measured below the cutoff}
        \State \textbf{continue}
      \EndIf
      \State $a_i \gets \textsc{AmplitudeEstimation}(\epsilon_{\textrm{meas} }, \ket{\psi}, \Pi_i)$ \Comment{Use the amplitude estimation algorithm, Algorithm~\ref{alg:final_algorithm} to estimate $a_i$}
    \EndFor
    \State \textbf{return} $\left\{ \Pi_j \right\}, \left\{ a_j \right\}$ \Comment{Return the estimated amplitude and basis vectors of corrections above the cutoff.}
  \end{algorithmic}
\end{algorithm}


In Section~\ref{sub:effect_of_noisy_measurements_on_the_qles_solver} we simulate the effect of applying different cutoff values, $\alpha$ to the measurements made within the matrix inversion loop of two CFD solvers, we find that the error in the final solution incurred by taking even very high values, $\alpha = 0.9$, are not significant compared to the errors when measuring the whole vector, given some $\epsilon_{\textrm{meas} }, \epsilon_{\textrm{inv} }$ parameters.


\section{Results}%
\label{sec:results}

We will now present two sets of results, firstly we will utilise the resource estimation of a matrix inversion algorithm for some useful block encodings given in~\cite{Sunderhauf2024a}.
We will also use the algorithm presented in~\cite{Lapworth2022b} to investigate the effect of reducing the precision, $\epsilon$ and number of peaks measured,  on the convergence of the classical part of the QLES algorithm, to reduce quantum resource requirements.

\subsection{Concrete Resource requirements} 
\label{sub:concrete_resource_requirements}

The CFD problem considered here is flow through a 1D channel.
This flow can be incompressible or compressible, and we can increase the accuracy of the CFD simulation by increasing the number of points, or stations, $s$, at which we measure.
The two system sizes we choose are 8 and 16 stations which, using the discretization scheme described in~\cite{Lapworth2022b}, produces a $2s \times 2s$ matrix.
This is the matrix which we will invert using QSVT.

We then provide a concrete resource estimation of a single call to the amplitude estimation oracle, $Q_\Pi$ of a QSVT based matrix inversion algorithm for these CFD matrices, and we include the cost of error correction on a fault tolerant quantum computer.
This oracle, as detailed above, requires two calls to the matrix inversion sub-routine and additional gates to implement the sign-dependent amplitude estimation routine.
We also consider what we call the naive case where no amplitude estimation is applied, but all qubits in the wavefunction are measured to produce an estimate of the absolute value of the state.
Using Chebyshev's inequality, where $a_i$ is the amplitude of each basis vector, and $\hat{a}_i$ is the estimator returned from the naive algorithm, $P(|\hat{a}_i - a_i | \geq \epsilon_{\textrm{meas} }) \leq \frac{\sigma(a_i)}{\epsilon_{\textrm{meas} }^2}$, implying we need $\mathcal{O}(\frac{1}{\epsilon_{\textrm{meas} }^2})$ shots.
However, as we measure all qubits simultaneously, we have an estimate of all amplitudes in these shots, as opposed to measuring $\hat{a}_i $ for each amplitude in turn using amplitude estimation.
We then use the value for $\langle Q_\Pi \rangle$ in Eqn.~\ref{eq:query_complexity_2} to compare the cost of using amplitude estimation with the naive scheme.

Table~\ref{tab:amplitude_encoding_resource_requrements} gives the cost of a single implementation of the matrix inversion algorithm, and the number of calls to the amplitude estimation oracle, given by Equation~\ref{eq:query_complexity_2}.
The assumptions made to calculate the physical qubit count and time taken are described in Appendix~\ref{app:resource_estimation_details}.

In this table we give the amount of error corrected resources required to run a single circuit, either as part of the amplitude estimation oracle or just the inversion circuit.
The columns of the table describe:
\begin{itemize}
  \item Stations is the number of stations in the CFD discretization, the matrix size is then $2s \times 2s$.
  \item The accuracy, $\epsilon_{\textrm{inv} }$ is a desired accuracy of the inversion, discussed in Section~\ref{sec:a_note_on_accuracy_epsilon}. $\epsilon_{\textrm{inv} }$ is split $90\%$ into the polynomial approximation of the $A^{-1}$ function, and $10\%$ into the accuracy of rotation gates \footnote{In a fault tolerant quantum computer arbitrary rotations must be decomposed into T + Clifford gates, more accurate rotations require more T gates.}. For simplicity, we set $\epsilon = \epsilon_{\textrm{meas} } = \epsilon_{\textrm{inv} }$.
  \item Phase factors is the number of terms required in QSVT to implement the polynomial approximating $A^{-1}$, the number of phase factors controls the overall length of the circuit as it requires one implementation of the block encoding and projector for each phase factor.
  \item $\kappa$ denotes the subnormalisation of the inversion circuit, $\kappa = \frac{2}{3 ||A||_{\infty} \textrm{min}(SV)}$ where $SV$ are the singular values of the matrix, and $3 ||A||_{\infty}$ is the subnormalisation of the block encoding, which is specific to the CFD matrices here~\cite{Sunderhauf2024a}. The factor $3||A||_{\infty}$ appears due to the restriction that the maximum value of any element in the matrix must be 1, and that there are three distinct diagonals. We also must divide by the smallest singular value as in the inversion we are approximating $\frac{1}{2\kappa'}A^{-1}$ with a polynomial over the range $[-1, 1] / [\frac{-1}{\kappa'}, \frac{1}{\kappa'}]$ so we do not need the polynomial to be a good approximation close to 0. Here we define the condition number, $\kappa'$ as the smallest singular value. This affects the number of times the circuit must be ran to get all flag qubits in the $\ket{0}$ state.
  \item Logical qubits are required for the algorithm and routing space, these require $\approx d^2$ physical qubits, where $d$ is the code distance required.
  \item Physical qubits is the total of the qubits in logical qubits plus an overhead for magic state distillation to produce T and Toffoli gates.
  \item T gates and Toffoli gates are reported separately, Toffoli gates implement the block encoding of the matrix, and T gates mostly implement the controlled phase factors.
  \item The time for a single oracle, $Q_\Pi$, assumes one magic state is consumed every $d$ code cycles.
  \item Number of oracle calls is calculated using Equation~\ref{eq:query_complexity} in the amplitude encoding case and $\frac{1}{\epsilon^2}$ in the naive case.
  \item Total time is given by oracle calls multiplied by the time for a single oracle call. \end{itemize}
In the final column, Percentage of Amplitudes, we pair results together for the same system with either measurement schemes, amplitude estimation or naive, and we calculate $N_{\textrm{amp} } = \lfloor t_{\textrm{naive} } / t_{\textrm{AE}} \rfloor$ as the number of amplitudes that can be measured using the amplitude estimation scheme in the time taken to complete a single naive measurement (which reports results for all amplitudes).
The percentage of amplitudes column then presents the lowest number of either $100\%$ or $N_{\textrm{amp} }$ as a percentage of the total number of amplitudes. 
This is  percentage of all amplitudes that can be measured one at a time using amplitude estimation in the time taken (i.e. $\frac{1}{\epsilon_{\textrm{meas} }^2}$ shots) to complete the naive scheme.

We see that in the small matrix sizes discussed here, the amplitude estimation oracle can be used to output the whole wavefunction in the time taken for the naive measurements, yet this will change as matrix sizes increase.
However, for the scenario considered here, where the output of the matrix inversion algorithm is a correction vector, we simulate a measurement scheme that requires the measurement of a very small number of the total basis vectors.
The qubit cost of the algorithm is the same as the naive version, as we have used the results of~\cite{Rossi2023a} to embed the matrix inversion QSVT routine in the amplitude estimation QSP routine, without using an addition al flag qubit.
Appendix~\ref{app:toeplitz_matrix_resource_estimation} presents the same costings for the Toeplitz matrix block encoding introduced in~\cite{Sunderhauf2024a}.

\begin{figure}[ht]
  \begin{center}
    \includegraphics[width=0.95\textwidth]{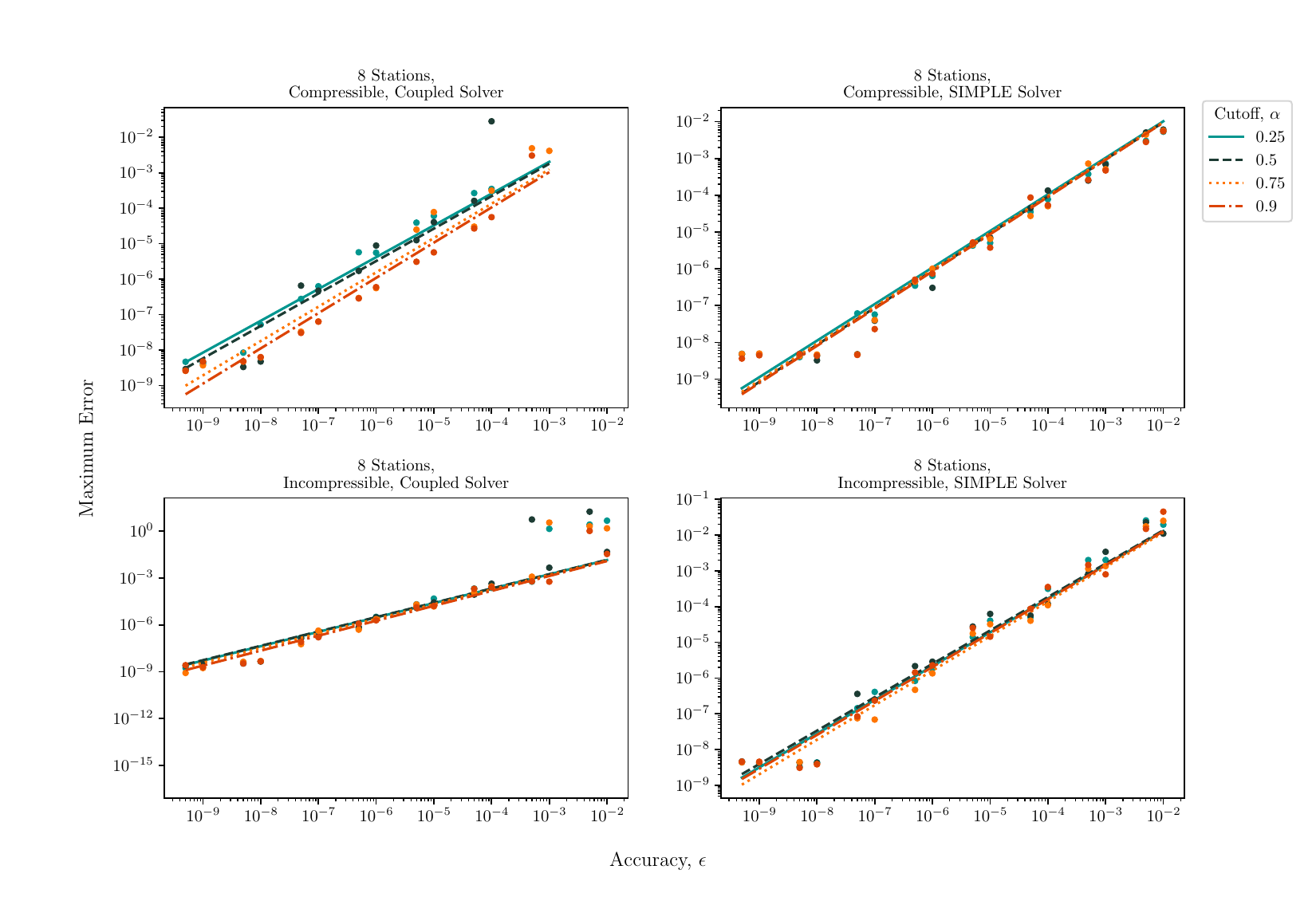}
  \end{center}
  \caption{Effect of increasing $\alpha$ and $\epsilon$ on the maximum error recorded for the solver of an $8 \times 8$ matrix for the four fluid cases considered here.}\label{fig:alpha_nozzle_chart_8}
\end{figure}
\begin{figure}
  \begin{center}
    \includegraphics[width=0.95\textwidth]{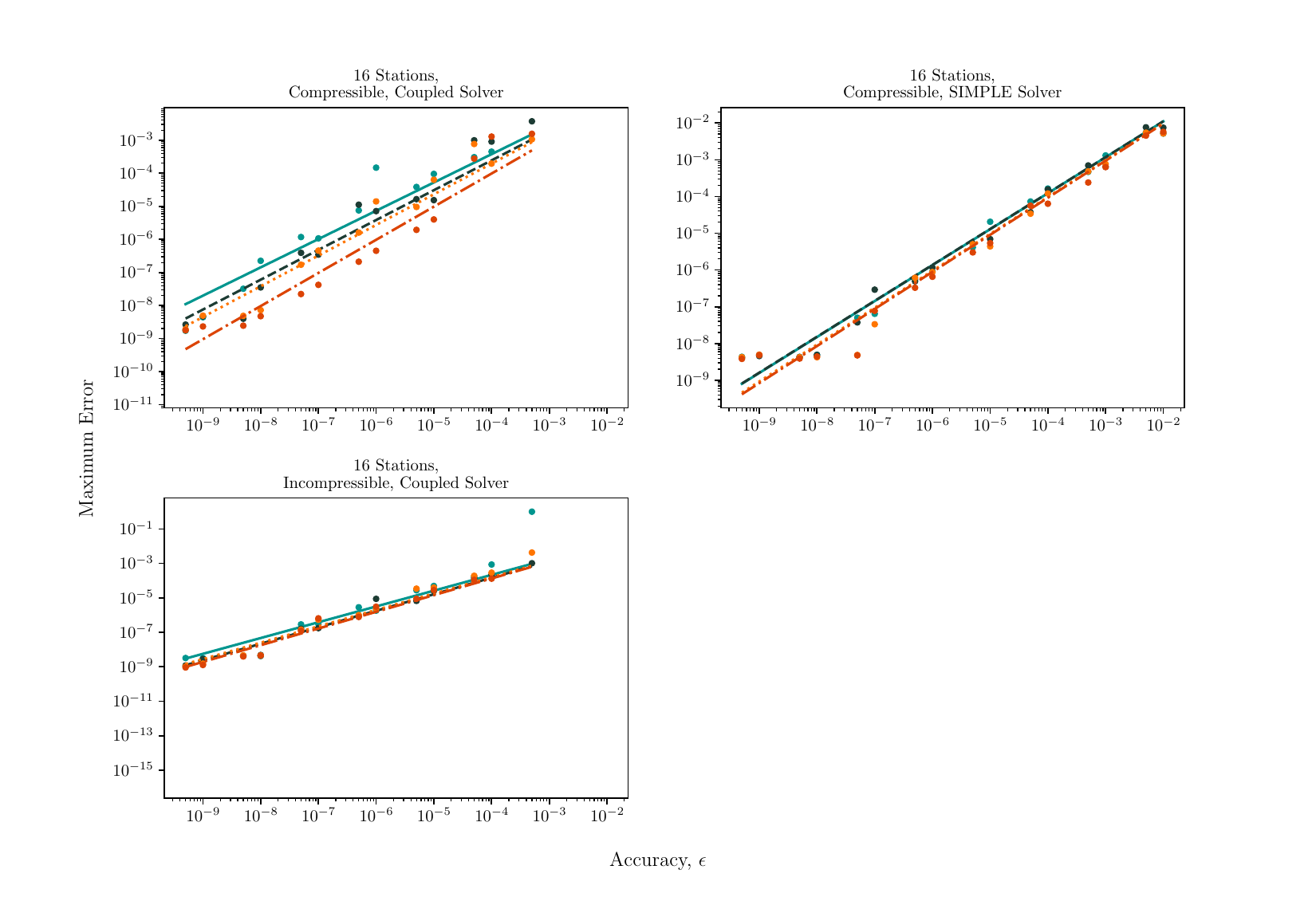}
  \end{center}
  \caption{Effect of increasing $\alpha$ and $\epsilon$ on the maximum error recorded for the solver of an $16 \times 16$ matrix for the compressible and incompressible fluid. The incompressible coupled solver has been omitted due to the time cost of obtaining enough unconverted results, many trials converged regardless of the value of $\alpha$.}\label{fig:alpha_nozzle_chart_16}
\end{figure}

\subsection{Effect of Noisy Measurements on the QLES Solver} 
\label{sub:effect_of_noisy_measurements_on_the_qles_solver}

We now need to test if our modifications to the classical CFD solver, i.e. using noisy measurements on a quantum computer and introducing the cutoff, have a detrimental effect on the overall scheme.
We simulate the effect of using a QSVT algorithm with accuracy $\epsilon_{\textrm{meas}}$ and cutoff $\alpha$ as a subroutine in a classical CFD solver, by running the outer loop of the CFD solver normally and modifying the corrections vector returned.
First, we characterise  the accuracy of the QSVT algorithm, $\epsilon_{\textrm{meas}}$ by applying a random Gaussian shift, described by a variance of $\epsilon_{\textrm{meas}}$ and mean of $0$ to the correction returned by the classical solver.
We then model the cutoff by  applying Equation~\ref{eq:alpha_cutoff} to the corrections vector.

The classical linear equation solver requires a tolerance $\epsilon_{\textrm{tol}}$, which stops the algorithm when the correction to at least one of the velocity, pressure, or compressible flow, $\rho$, are below $\epsilon_{\textrm{tol}}$.
In preliminary investigations, we find that, without introducing an $\alpha$-cutoff (alternately $\alpha = 0$), most cases of the solver converge if  $\epsilon_{\textrm{tol} } = \epsilon_{\textrm{meas}}$.
We are interested in the cases where $\epsilon_{\textrm{meas}} > \epsilon_{\textrm{tol}}, \alpha > 0$.
For simplicity, we now only discuss the maximum correction (or error) returned over the three parameters of interest for the final correction vector.
The final correction vector is returned when either the tolerance condition of the classical solver is met, or we have exceeded some number of total iterations, which for the results here is $10^8$ \footnote{Other preliminary investigations used $10^{10}$ as the total iterations, but there were no instances that converged below $10^8$ iterations.}.
We then set the tolerance of the classical algorithm to $1 \times 10^{-9}$ and investigate the effect of increasing $\epsilon, \alpha$ on the maximum error.

In Figures~\ref{fig:alpha_nozzle_chart_8} and ~\ref{fig:alpha_nozzle_chart_16} we show the effect of accuracy, $\epsilon_{\textrm{meas}}$, and cutoff value, $\alpha$, on the maximum error recorded for $s = 8, 16$ examples.
Both accuracy and maximum error have been plotted on a log scale, and a linear fit has been drawn.
We see that the maximum error in the correction vector rises when the accuracy is decreased, as can be expected.
More interestingly, increasing the value of $\alpha$, i.e. measuring fewer amplitudes, seems to have a small effect on the final result.
This means we can perform fewer measurements overall by setting $\alpha$ to high values, e.g. $\alpha = 0.9$ for a moderate sacrifice in the final maximum error.
The coupled solutions in Figure~\ref{fig:alpha_nozzle_chart_8} show some outliers for $\epsilon>10^{-3}$. These are cases where a low-precision feedback loop caused the non-linear
solver to fail to converge \cite{Lapworth2022c}.
All other cases converged to the desired accuracy.

\begin{turnpage}
  \begin{table}[ht]
  \rowcolors{2}{gray!20}{white}
  \begin{center}
    \small
  \begin{tblr}{colspec={p{1.6cm}p{1.45cm}p{1.45cm}p{1.45cm}p{1.45cm}p{1.45cm}p{1.45cm}p{1.45cm}p{1.45cm}p{1.45cm}p{1.45cm}p{1.45cm}p{1.45cm}},
row{odd}={bg=lightgray},
row{1}={bg=white}}
Amplitude Estimation & Accuracy & Stations & Phase Factors & $\kappa$ &  Logical Qubits &  Physical Qubits & T Gates & Toffoli Gates & Oracle Time (s) &  Oracle Calls & Total Time (Days) & Percentage Of Amplitudes \\
\hline 
True & $1\times 10^{-5}$ & $8$ & $8.83\times 10^{4}$ & 48.13 & 35 & $4.23\times 10^{4}$ & $7.21\times 10^{8}$ & $1.59\times 10^{6}$ & $8.73\times 10^{3}$ & $1.75\times 10^{6}$ & $1.77\times 10^{5}$ & - \\
False & $1\times 10^{-5}$ & $8$ & $8.83\times 10^{4}$ & 48.13 & 35 & $5.69\times 10^{4}$ & $1.78\times 10^{8}$ & $7.95\times 10^{5}$ & $1.99\times 10^{3}$ & $1.00\times 10^{10}$ & $2.30\times 10^{8}$ & 100 \\
True & $1\times 10^{-4}$ & $8$ & $7.68\times 10^{4}$ & 48.13 & 35 & $4.23\times 10^{4}$ & $5.83\times 10^{8}$ & $1.38\times 10^{6}$ & $7.07\times 10^{3}$ & $1.64\times 10^{5}$ & $1.34\times 10^{4}$ & - \\
False & $1\times 10^{-4}$ & $8$ & $7.68\times 10^{4}$ & 48.13 & 35 & $5.69\times 10^{4}$ & $1.43\times 10^{8}$ & $6.91\times 10^{5}$ & $1.61\times 10^{3}$ & $1.00\times 10^{8}$ & $1.86\times 10^{6}$ & 100 \\
True & $1\times 10^{-3}$ & $8$ & $6.53\times 10^{4}$ & 48.13 & 35 & $4.23\times 10^{4}$ & $4.58\times 10^{8}$ & $1.17\times 10^{6}$ & $5.56\times 10^{3}$ & $1.50\times 10^{4}$ & 963 & - \\
False & $1\times 10^{-3}$ & $8$ & $6.53\times 10^{4}$ & 48.13 & 35 & $5.69\times 10^{4}$ & $1.13\times 10^{8}$ & $5.87\times 10^{5}$ & $1.26\times 10^{3}$ & $1.00\times 10^{6}$ & $1.46\times 10^{4}$ & 93.75 \\
True & 0.01 & $8$ & $5.37\times 10^{4}$ & 48.13 & 35 & $4.23\times 10^{4}$ & $3.46\times 10^{8}$ & $9.66\times 10^{5}$ & $4.20\times 10^{3}$ & $1.30\times 10^{3}$ & 63 & - \\
False & 0.01 & $8$ & $5.37\times 10^{4}$ & 48.13 & 35 & $5.69\times 10^{4}$ & $8.48\times 10^{7}$ & $4.83\times 10^{5}$ & 954.20 & $1.00\times 10^{4}$ & 111 & 6.25 \\
True & $1\times 10^{-5}$ & $16$ & $2.57\times 10^{5}$ & 131.05 & 41 & $4.61\times 10^{4}$ & $4.20\times 10^{9}$ & $5.65\times 10^{6}$ & $5.49\times 10^{4}$ & $1.75\times 10^{6}$ & $1.11\times 10^{6}$ & - \\
False & $1\times 10^{-5}$ & $16$ & $2.57\times 10^{5}$ & 131.05 & 41 & $4.41\times 10^{4}$ & $1.03\times 10^{9}$ & $2.82\times 10^{6}$ & $1.25\times 10^{4}$ & $1.00\times 10^{10}$ & $1.45\times 10^{9}$ & 100 \\
True & $1\times 10^{-4}$ & $16$ & $2.25\times 10^{5}$ & 131.05 & 41 & $4.61\times 10^{4}$ & $3.44\times 10^{9}$ & $4.95\times 10^{6}$ & $4.49\times 10^{4}$ & $1.64\times 10^{5}$ & $8.52\times 10^{4}$ & - \\
False & $1\times 10^{-4}$ & $16$ & $2.25\times 10^{5}$ & 131.05 & 41 & $4.41\times 10^{4}$ & $8.44\times 10^{8}$ & $2.47\times 10^{6}$ & $1.02\times 10^{4}$ & $1.00\times 10^{8}$ & $1.19\times 10^{7}$ & 100 \\
True & $1\times 10^{-3}$ & $16$ & $1.93\times 10^{5}$ & 131.05 & 41 & $4.61\times 10^{4}$ & $2.74\times 10^{9}$ & $4.25\times 10^{6}$ & $3.59\times 10^{4}$ & $1.50\times 10^{4}$ & $6.21\times 10^{3}$ & - \\
False & $1\times 10^{-3}$ & $16$ & $1.93\times 10^{5}$ & 131.05 & 41 & $4.41\times 10^{4}$ & $6.72\times 10^{8}$ & $2.13\times 10^{6}$ & $8.17\times 10^{3}$ & $1.00\times 10^{6}$ & $9.46\times 10^{4}$ & 46.88 \\
True & 0.01 & $16$ & $1.62\times 10^{5}$ & 131.05 & 41 & $4.61\times 10^{4}$ & $2.11\times 10^{9}$ & $3.55\times 10^{6}$ & $2.77\times 10^{4}$ & $1.30\times 10^{3}$ & 415 & - \\
False & 0.01 & $16$ & $1.62\times 10^{5}$ & 131.05 & 41 & $4.41\times 10^{4}$ & $5.17\times 10^{8}$ & $1.78\times 10^{6}$ & $6.29\times 10^{3}$ & $1.00\times 10^{4}$ & 729 & 3.12 \\
\end{tblr}

  \end{center}
  \caption{
    Resource requirements for the uncoupled compressible fluid in a nozzle scenario.
    Number of phase factors characterises how many phase factors are required to approximate $A^{-1}$ at accuracy $\epsilon$, $\kappa$ is the subnormalisation of the matrix to be inverted.
    Oracle time is the time taken to implement an amplitude encoding oracle where Amplitude Estimation is True, and a single QSVT circuit otherwise.
    Oracle calls is given by Equation~\ref{eq:query_complexity} when Amplitude Estimation is True, and $\frac{1}{\epsilon^2}$ otherwise.
    Total time is oracle calls multiplied by oracle time.
    Percentage of amplitudes characterises the percentage of the basis vectors that can be measured using Amplitude Estimation in the time taken for the naive case.
  }\label{tab:amplitude_encoding_resource_requrements}
\end{table}
\end{turnpage}

\section{Discussion}%
\label{sec:discussion}

We have studied the use of amplitude estimation as a method of measuring outputs of a QSVT algorithm, specifically for CFD use cases.
We have adapted the algorithm presented in~\cite{Rall2023} for use in a setting where the implemented unitary is itself a QSVT operator, and we have used the estimates given in~\cite{Rall2023} to estimate the number of times an operator needs to be measured to estimate a single correction peak from a CFD calculation.
This allows us to estimate physical resources for a single amplitude.

The time resources required by this algorithm are daunting, especially when considering the fact that this is the inner loop of a CFD calculation.
The idea introduced here, of using cut-off $\alpha$ reduces the number of times the algorithm will be used, but not the resources for a single implementation of matrix inversion.
The number of phase factors required to implement the matrix inversion polynomial is the dominant factor in these circuits, reducing this number whilst maintaining $\epsilon_{inv}$ is the subject of future work.
As the function approximating $A^{-1}$ is real, an antisymmetric list of phase factors is guaranteed to exist.
However, whilst there has been recent advances in finding phase factors for polynomials~\cite{Berntson2024, Alexis2024b}, this is not for an antisymmetric list.
Work into constructing antisymmetric lists of phase factors will also be constructive.

\pagebreak
\section*{Acknowledgements}
We would like to thank Christoph S\"underhauf and Bjorn Berntson for helpful discussions and reviewing this manuscript. This work was partially funded by Innovate UK, grant number 10071684.

\bibliography{bibliography.bib}

\begin{thebibliography}{10}
\providecommand{\url}[1]{\texttt{#1}}
\providecommand{\urlprefix}{URL }
\expandafter\ifx\csname urlstyle\endcsname\relax
  \providecommand{\doi}[1]{doi:\discretionary{}{}{}#1}\else
  \providecommand{\doi}{doi:\discretionary{}{}{}\begingroup \urlstyle{rm}\Url}\fi
\providecommand{\eprint}[2][]{\url{#2}}

\bibitem{Gilyen2019}
A.~Gily{\'e}n, Y.~Su, G.~H. Low and N.~Wiebe,
\newblock \emph{Quantum singular value transformation and beyond: {{Exponential}} improvements for quantum matrix arithmetics},
\newblock In \emph{Proceedings of the 51st {{Annual ACM SIGACT Symposium}} on {{Theory}} of {{Computing}}}, pp. 193--204. ACM, Phoenix AZ USA,
\newblock ISBN 978-1-4503-6705-9,
\newblock \doi{10.1145/3313276.3316366} (2019).

\bibitem{Rall2023}
P.~Rall and B.~Fuller,
\newblock \emph{Amplitude {{Estimation}} from {{Quantum Signal Processing}}},
\newblock Quantum \textbf{7}, 937 (2023),
\newblock \doi{10.22331/q-2023-03-02-937}.

\bibitem{Shor1994}
P.~W. Shor,
\newblock \emph{Polynomial-{{Time Algorithms}} for {{Prime Factorization}} and {{Discrete Logarithms}} on a {{Quantum Computer}}},
\newblock In \emph{{{AT}}\&{{T Research}}}, pp. 20--22. IEEE Computer Society Press (1994).

\bibitem{Lapworth2022b}
L.~Lapworth,
\newblock \emph{A {{Hybrid Quantum-Classical CFD Methodology}} with {{Benchmark HHL Solutions}}} (2022), \eprint{2206.00419}.

\bibitem{James2001}
D.~F.~V. James, P.~G. Kwiat, W.~J. Munro and A.~G. White,
\newblock \emph{On the {{Measurement}} of {{Qubits}}},
\newblock Physical Review A \textbf{64}(5), 052312 (2001),
\newblock \doi{10.1103/PhysRevA.64.052312},
\newblock \eprint{quant-ph/0103121}.

\bibitem{Hradil1997}
Z.~Hradil,
\newblock \emph{Quantum-state estimation},
\newblock Physical Review A \textbf{55}(3), R1561 (1997),
\newblock \doi{10.1103/PhysRevA.55.R1561}.

\bibitem{Torlai2018}
G.~Torlai, G.~Mazzola, J.~Carrasquilla, M.~Troyer, R.~Melko and G.~Carleo,
\newblock \emph{Many-body quantum state tomography with neural networks},
\newblock Nature Physics \textbf{14}(5), 447 (2018),
\newblock \doi{10.1038/s41567-018-0048-5},
\newblock \eprint{1703.05334}.

\bibitem{Cramer2010}
M.~Cramer, M.~B. Plenio, S.~T. Flammia, R.~Somma, D.~Gross, S.~D. Bartlett, O.~{Landon-Cardinal}, D.~Poulin and Y.-K. Liu,
\newblock \emph{Efficient quantum state tomography},
\newblock Nature Communications \textbf{1}(1), 149 (2010),
\newblock \doi{10.1038/ncomms1147}.

\bibitem{Rossi2023a}
Z.~M. Rossi, J.~L. Ceroni and I.~L. Chuang,
\newblock \emph{Modular quantum signal processing in many variables},
\newblock \doi{10.48550/arXiv.2309.16665} (2023), \eprint{2309.16665}.

\bibitem{Manzano2023}
A.~Manzano, D.~Musso and A.~Leitao,
\newblock \emph{Real quantum amplitude estimation},
\newblock EPJ Quantum Technology \textbf{10}(1), 1 (2023),
\newblock \doi{10.1140/epjqt/s40507-023-00159-0}.

\bibitem{Brassard2002}
G.~Brassard, P.~Hoyer, M.~Mosca and A.~Tapp,
\newblock \emph{Quantum {{Amplitude Amplification}} and {{Estimation}}},
\newblock \doi{10.1090/conm/305/05215} (2002), \eprint{quant-ph/0005055}.

\bibitem{Grover2001}
L.~K. Grover,
\newblock \emph{From {{Schr{\"o}dinger}}'s equation to the quantum search algorithm},
\newblock Pramana - Journal of Physics \textbf{56}(2-3), 333 (2001),
\newblock \doi{10.1119/1.1359518}.

\bibitem{Kitaev1996}
A.~Kitaev,
\newblock \emph{Quantum measurements and the {{Abelian Stabilizer Problem}}},
\newblock Tech. Rep. TR96-003, Electronic Colloquium on Computational Complexity (ECCC) (1996).

\bibitem{Aaronson2019}
S.~Aaronson and P.~Rall,
\newblock \emph{Quantum {{Approximate Counting}}, {{Simplified}}},
\newblock In \emph{2020 {{Symposium}} on {{Simplicity}} in {{Algorithms}} ({{SOSA}})}, Proceedings, pp. 24--32. {Society for Industrial and Applied Mathematics},
\newblock \doi{10.1137/1.9781611976014.5} (2019).

\bibitem{Suzuki2020}
Y.~Suzuki, S.~Uno, R.~Raymond, T.~Tanaka, T.~Onodera and N.~Yamamoto,
\newblock \emph{Amplitude estimation without phase estimation},
\newblock Quantum Information Processing \textbf{19}(2), 75 (2020),
\newblock \doi{10.1007/s11128-019-2565-2}.

\bibitem{Wie2019}
C.-R. Wie,
\newblock \emph{Simpler quantum counting},
\newblock Quantum Information and Computation \textbf{19}(11\&12), 967 (2019),
\newblock \doi{10.26421/QIC19.11-12-5}.

\bibitem{Dobsicek2007a}
M.~Dobsicek, G.~Johansson, V.~S. Shumeiko and G.~Wendin,
\newblock \emph{Arbitrary accuracy iterative phase estimation algorithm as a two qubit benchmark},
\newblock Physical Review A \textbf{76}(3), 030306 (2007),
\newblock \doi{10.1103/PhysRevA.76.030306},
\newblock \eprint{quant-ph/0610214}.

\bibitem{Grinko2021a}
D.~Grinko, J.~Gacon, C.~Zoufal and S.~Woerner,
\newblock \emph{Iterative quantum amplitude estimation},
\newblock npj Quantum Information \textbf{7}(1), 1 (2021),
\newblock \doi{10.1038/s41534-021-00379-1}.

\bibitem{Callison2023}
A.~Callison and D.~E. Browne,
\newblock \emph{Improved maximum-likelihood quantum amplitude estimation},
\newblock \doi{10.48550/arXiv.2209.03321} (2023), \eprint{2209.03321}.

\bibitem{Mizuta2023a}
K.~Mizuta and K.~Fujii,
\newblock \emph{Recursive {{Quantum Eigenvalue}}/{{Singular-Value Transformation}}: {{Analytic Construction}} of {{Matrix Sign Function}} by {{Newton Iteration}}},
\newblock \doi{10.48550/arXiv.2304.13330} (2023), \eprint{2304.13330}.

\bibitem{Rossi2023b}
Z.~M. Rossi and I.~L. Chuang,
\newblock \emph{Semantic embedding for quantum algorithms},
\newblock \doi{10.48550/arXiv.2304.14392} (2023), \eprint{2304.14392}.

\bibitem{Clopper1934}
C.~J. Clopper and E.~S. Pearson,
\newblock \emph{The {{Use}} of {{Confidence}} or {{Fiducial Limits Illustrated}} in the {{Case}} of the {{Binomial}}},
\newblock Biometrika \textbf{26}(4), 404 (1934),
\newblock \doi{10.1093/biomet/26.4.404}.

\bibitem{Shende2006a}
V.~V. Shende, S.~S. Bullock and I.~L. Markov,
\newblock \emph{Synthesis of {{Quantum Logic Circuits}}},
\newblock IEEE Transactions on Computer-Aided Design of Integrated Circuits and Systems \textbf{25}(6), 1000 (2006),
\newblock \doi{10.1109/TCAD.2005.855930},
\newblock \eprint{quant-ph/0406176}.

\bibitem{Hills2007}
N.~Hills,
\newblock \emph{Achieving high parallel performance for an unstructured unsteady turbomachinery {{CFD}} code},
\newblock The Aeronautical Journal \textbf{111}(1117), 185 (2007),
\newblock \doi{10.1017/S0001924000004449}.

\bibitem{Lapworth2022c}
L.~Lapworth,
\newblock \emph{Implicit {{Hybrid Quantum-Classical CFD Calculations}} using the {{HHL Algorithm}}},
\newblock \doi{10.48550/arXiv.2209.07964} (2022), \eprint{2209.07964}.

\bibitem{Sunderhauf2024a}
C.~S{\"u}nderhauf, E.~Campbell and J.~Camps,
\newblock \emph{Block-encoding structured matrices for data input in quantum computing},
\newblock Quantum \textbf{8}, 1226 (2024),
\newblock \doi{10.22331/q-2024-01-11-1226}.

\bibitem{Berntson2024}
B.~K. Berntson and C.~S{\"u}nderhauf,
\newblock \emph{Complementary polynomials in quantum signal processing},
\newblock \doi{10.48550/arXiv.2406.04246} (2024), \eprint{2406.04246}.

\bibitem{Alexis2024b}
M.~Alexis, L.~Lin, G.~Mnatsakanyan, C.~Thiele and J.~Wang,
\newblock \emph{Infinite quantum signal processing for arbitrary {{Szeg{\H o}}} functions}  (2024),
\newblock \doi{10.48550/arXiv.2407.05634},
\newblock \eprint{2407.05634}.

\bibitem{Litinski2019}
D.~Litinski,
\newblock \emph{A {{Game}} of {{Surface Codes}}: {{Large-Scale Quantum Computing}} with {{Lattice Surgery}}},
\newblock Quantum \textbf{3}, 128 (2019),
\newblock \doi{10.22331/q-2019-03-05-128}.

\bibitem{Litinski2019d}
D.~Litinski,
\newblock \emph{Magic {{State Distillation}}: {{Not}} as {{Costly}} as {{You Think}}},
\newblock arXiv:1905.06903 [quant-ph]  (2019),
\newblock \eprint{1905.06903}.

\bibitem{Fowler2012}
A.~G. Fowler, M.~Mariantoni, J.~M. Martinis and A.~N. Cleland,
\newblock \emph{Surface codes: {{Towards}} practical large-scale quantum computation},
\newblock Physical Review A - Atomic, Molecular, and Optical Physics \textbf{86}(3) (2012),
\newblock \doi{10.1103/PhysRevA.86.032324}.

\bibitem{Acharya2023a}
R.~Acharya, I.~Aleiner, R.~Allen, T.~I. Andersen, M.~Ansmann, F.~Arute, K.~Arya, A.~Asfaw, J.~Atalaya, R.~Babbush, D.~Bacon, J.~C. Bardin \emph{et~al.},
\newblock \emph{Suppressing quantum errors by scaling a surface code logical qubit},
\newblock Nature \textbf{614}(7949), 676 (2023),
\newblock \doi{10.1038/s41586-022-05434-1}.

\bibitem{Krinner2022b}
S.~Krinner, N.~Lacroix, A.~Remm, A.~Di~Paolo, E.~Genois, C.~Leroux, C.~Hellings, S.~Lazar, F.~Swiadek, J.~Herrmann, G.~J. Norris, C.~K. Andersen \emph{et~al.},
\newblock \emph{Realizing repeated quantum error correction in a distance-three surface code},
\newblock Nature \textbf{605}(7911), 669 (2022),
\newblock \doi{10.1038/s41586-022-04566-8}.

\end{thebibliography}

\pagebreak
\appendix
\section{Resource Estimation Details}
\label{app:resource_estimation_details} 
To produce the concrete resource requirements we must calculate the additional resources required for error correction once we have calculated the logical qubit and non-Clifford gate count.
We must make some assumptions about the device to use, which is based on the surface code on a 2D grid, following the scheme given in ``A Game of Surface Codes"~\cite{Litinski2019}.

We choose a target failure probability of $p_\text{fail}=1\%$ for a full execution of the quantum algorithm, which we divide into error budgets $p_\text{fail}^\text{log}=0.9\%$ for logical errors, and $p_\text{fail}^\text{MSD}=0.1\%$ for undetected errors in magic state distillation.

For the systems considered here, we use a two level magic state factory with asymmetric code distances, based on the scheme laid out in~\cite{Litinski2019d}.
These factories produce high quality $T$ states using smaller magic state factories to input some lower quality distilled $T$ states.
The longer circuits seen here use the (15-to-1)${}^4_{9,3,3}$ $\times$ (20-to-4)${}_{15,7,9}$ factory from \cite{Litinski2019d}. 
It has a sufficiently low failure probability, below the target error $p_\text{fail}^\text{MSD}/N_\text{T}$
A smaller code distance than $d$ in the logical computation can be used for the magic state factory to reduce its footprint and runtime. In fact, rectangular code patches with distinct distances for X, Z, and time (as indicated by the subscripts) can be used as the factory is more prone to Z errors than X.

Using the~\cite{Litinski2019} scheme, we assume the computation proceeds as fast as consuming one magic state qubit per logical clock cycle, where a logical clock cycle is equivalent to $d$ code cycles, and $d$ is the code distance.
Consequently, we ensure that the number of magic state factories available is high enough that a single magic state is available every logical cycle, which typically requires multiple magic state factories.
The length of the computation is $N_\text{T}$ logical cycles.

The logical error budget bounds the allowed logical failure probability per logical cycle, which is given by the Fowler-Devitt-Jones formula~\cite{Fowler2012}. Hence the computational code distance $d$ must be chosen such that
\begin{equation}
A \left (\frac{p}{p_{\textrm{thr}}} \right )^{\frac{d + 1}{2}} \le \frac{p_\text{fail}^\text{log}}{N_\text{T} N_{\textrm{qubits}}} ,
  \label{eq:fowler_formula}
\end{equation}
where $p$ is the probability of a physical error, $p_\text{thr}\approx0.01$ is the threshold of the surface code, and $A=0.1$ is a numerically determined constant.

In order to estimate the physical resources, we model a 2D superconducting device, with an error rate one order of magnitude better than current superconducting devices~\cite{Acharya2023a, Krinner2022b}, i.e. $p  = 0.01 \%$.
This allows us to solve Equation~\ref{eq:fowler_formula} for $d$.
The total number of logical qubits $n_{L}$ is given by the number of qubits required by the algorithm and the routing required by the \textit{fast-block} layout~\cite[Figure 13]{Litinski2019}.
The total number of physical qubits is then $(2d^2 -1)n_L$ for algorithm and routing in the rotated surface code, together with those required by the magic state factories.

\section{Modelling $a_{\textrm{max}}$}
\label{app:modelling_a_max} 

\begin{figure}
  \begin{center}
    \includegraphics[width=0.95\textwidth]{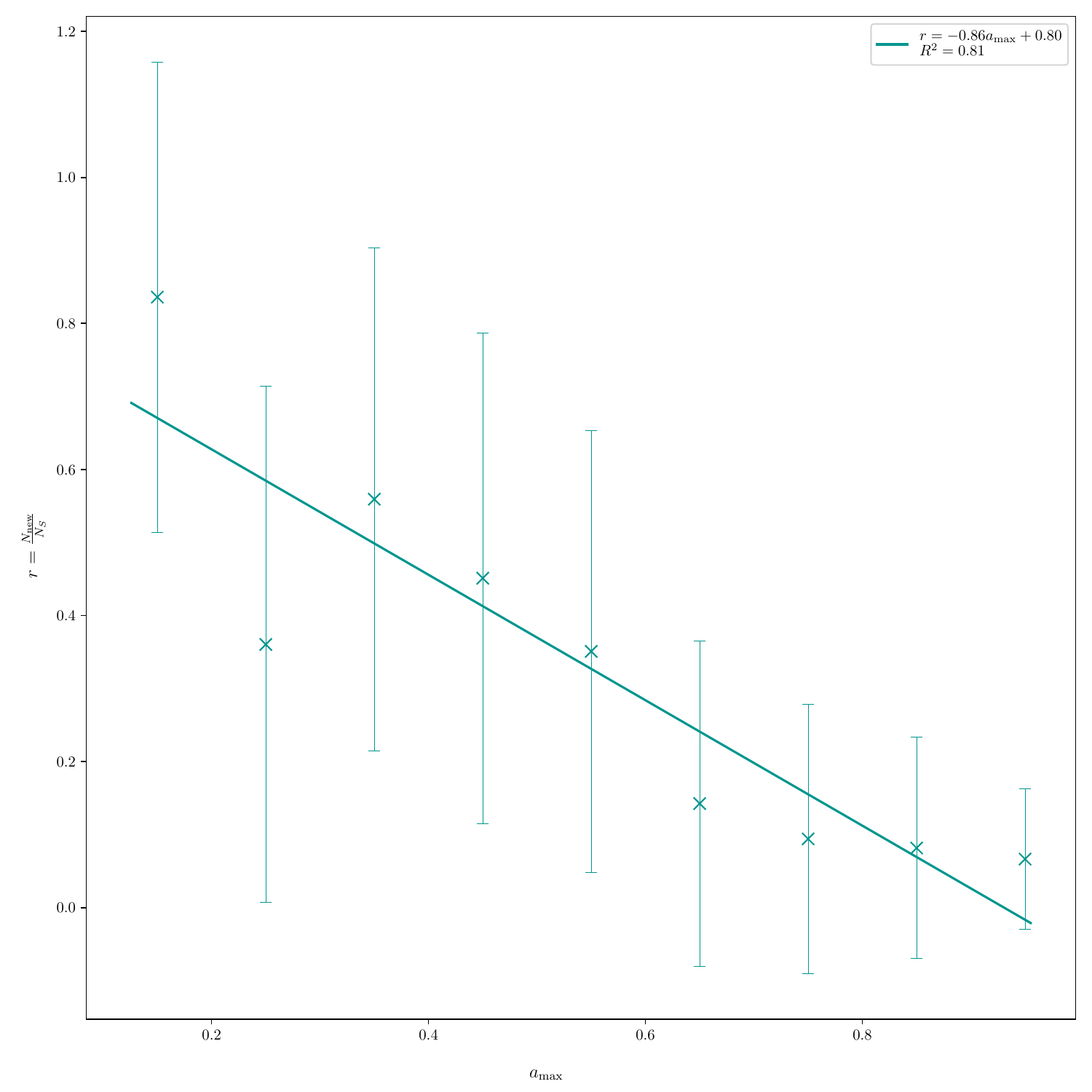}
  \end{center}
  \caption{Linear fit of $r$ against $a_{\textrm{max} }$ which can be used to give an estimate of $a_{\textrm{max} }$ in the `burn-in' period.}\label{fig:a_max_r_plot_fit}
\end{figure}

To model $a_{\textrm{max}}$ we ran the CFD solver over the incompressible and compressible fluid cases for 3 - 8 qubit systems and saved all output vectors.
We then simulate measurement of these wavefunctions by sampling for $N_{\textrm{shots}} \in [10, 50, 100, 500, 1 \times 10^{3}, 5 \times 10^{3}, 1 \times 10^{4}, 5 \times 10^{4}, 1 \times 10^{5}]$ and recording the basis vector measured.
We follow the algorithm detailed in Section~\ref{sub:modelling_a_max}, recording the number of unique basis vectors seen, $N_{\textrm{new} }$ and the ratio $r = \frac{N_{\textrm{new} }}{N_S}$.
The $a_{\textrm{max} }$ data was separated into decile bins, and the ratio $r$ was averaged for these bins.
A linear fit was calculated for the averaged $r$, which is shown in Figure~\ref{fig:a_max_r_plot_fit}.
The large error bars represent one standard deviation of the $r$ values, which whilst large, still show the required relationship when averaged.
As this relationship is used as a heuristic to identify the largest peaks, we argue that this loose relationship suffices for out purposes.

The fit is:
\begin{equation}
  r = -0.86 a_{\textrm{max} }  + 0.80.
  \label{eq:a_max_r_app}
\end{equation}
This can be solved for some real instance of $r$ to give an estimate of $a_{\textrm{max} }$.


\section{Toeplitz Matrix Resource Estimation} 
\label{app:toeplitz_matrix_resource_estimation}

In \cite{Sunderhauf2024a} a block encoding for the Toeplitz matrix, which has a wide number of potential applications, was introduced.
Here we apply the same resource estimation used for the compressible and incompressible nozzle in the main text to the Toeplitz matrix.

\begin{turnpage}
  \begin{table}[p]
  \rowcolors{2}{gray!20}{white}
  \begin{center}
    \small
  \begin{tblr}{colspec={p{1.6cm}p{1.45cm}p{1.45cm}p{1.45cm}p{1.45cm}p{1.45cm}p{1.45cm}p{1.45cm}p{1.45cm}p{1.45cm}p{1.45cm}p{1.45cm}p{1.45cm}},
row{odd}={bg=lightgray},
row{1}={bg=white}}
Amplitude Estimation & Accuracy & Stations & Phase Factors & $\kappa$ &  Logical Qubits &  Physical Qubits & T Gates & Toffoli Gates & Oracle Time (s) &  Oracle Calls & Total Time (Days) & Percentage Of Amplitudes \\
\hline 
True & $1\times 10^{-5}$ & $8$ & $8.83\times 10^{4}$ & 48.13 & 35 & $4.23\times 10^{4}$ & $7.21\times 10^{8}$ & $1.59\times 10^{6}$ & $8.73\times 10^{3}$ & $1.75\times 10^{6}$ & $1.77\times 10^{5}$ & - \\
False & $1\times 10^{-5}$ & $8$ & $8.83\times 10^{4}$ & 48.13 & 35 & $5.69\times 10^{4}$ & $1.78\times 10^{8}$ & $7.95\times 10^{5}$ & $1.99\times 10^{3}$ & $1.00\times 10^{10}$ & $2.30\times 10^{8}$ & 100 \\
True & $1\times 10^{-4}$ & $8$ & $7.68\times 10^{4}$ & 48.13 & 35 & $4.23\times 10^{4}$ & $5.83\times 10^{8}$ & $1.38\times 10^{6}$ & $7.07\times 10^{3}$ & $1.64\times 10^{5}$ & $1.34\times 10^{4}$ & - \\
False & $1\times 10^{-4}$ & $8$ & $7.68\times 10^{4}$ & 48.13 & 35 & $5.69\times 10^{4}$ & $1.43\times 10^{8}$ & $6.91\times 10^{5}$ & $1.61\times 10^{3}$ & $1.00\times 10^{8}$ & $1.86\times 10^{6}$ & 100 \\
True & $1\times 10^{-3}$ & $8$ & $6.53\times 10^{4}$ & 48.13 & 35 & $4.23\times 10^{4}$ & $4.58\times 10^{8}$ & $1.17\times 10^{6}$ & $5.56\times 10^{3}$ & $1.50\times 10^{4}$ & 963 & - \\
False & $1\times 10^{-3}$ & $8$ & $6.53\times 10^{4}$ & 48.13 & 35 & $5.69\times 10^{4}$ & $1.13\times 10^{8}$ & $5.87\times 10^{5}$ & $1.26\times 10^{3}$ & $1.00\times 10^{6}$ & $1.46\times 10^{4}$ & 93.75 \\
True & 0.01 & $8$ & $5.37\times 10^{4}$ & 48.13 & 35 & $4.23\times 10^{4}$ & $3.46\times 10^{8}$ & $9.66\times 10^{5}$ & $4.20\times 10^{3}$ & $1.30\times 10^{3}$ & 63 & - \\
False & 0.01 & $8$ & $5.37\times 10^{4}$ & 48.13 & 35 & $5.69\times 10^{4}$ & $8.48\times 10^{7}$ & $4.83\times 10^{5}$ & 954.20 & $1.00\times 10^{4}$ & 111 & 6.25 \\
True & $1\times 10^{-5}$ & $16$ & $2.57\times 10^{5}$ & 131.05 & 41 & $4.61\times 10^{4}$ & $4.20\times 10^{9}$ & $5.65\times 10^{6}$ & $5.49\times 10^{4}$ & $1.75\times 10^{6}$ & $1.11\times 10^{6}$ & - \\
False & $1\times 10^{-5}$ & $16$ & $2.57\times 10^{5}$ & 131.05 & 41 & $4.41\times 10^{4}$ & $1.03\times 10^{9}$ & $2.82\times 10^{6}$ & $1.25\times 10^{4}$ & $1.00\times 10^{10}$ & $1.45\times 10^{9}$ & 100 \\
True & $1\times 10^{-4}$ & $16$ & $2.25\times 10^{5}$ & 131.05 & 41 & $4.61\times 10^{4}$ & $3.44\times 10^{9}$ & $4.95\times 10^{6}$ & $4.49\times 10^{4}$ & $1.64\times 10^{5}$ & $8.52\times 10^{4}$ & - \\
False & $1\times 10^{-4}$ & $16$ & $2.25\times 10^{5}$ & 131.05 & 41 & $4.41\times 10^{4}$ & $8.44\times 10^{8}$ & $2.47\times 10^{6}$ & $1.02\times 10^{4}$ & $1.00\times 10^{8}$ & $1.19\times 10^{7}$ & 100 \\
True & $1\times 10^{-3}$ & $16$ & $1.93\times 10^{5}$ & 131.05 & 41 & $4.61\times 10^{4}$ & $2.74\times 10^{9}$ & $4.25\times 10^{6}$ & $3.59\times 10^{4}$ & $1.50\times 10^{4}$ & $6.21\times 10^{3}$ & - \\
False & $1\times 10^{-3}$ & $16$ & $1.93\times 10^{5}$ & 131.05 & 41 & $4.41\times 10^{4}$ & $6.72\times 10^{8}$ & $2.13\times 10^{6}$ & $8.17\times 10^{3}$ & $1.00\times 10^{6}$ & $9.46\times 10^{4}$ & 46.88 \\
True & 0.01 & $16$ & $1.62\times 10^{5}$ & 131.05 & 41 & $4.61\times 10^{4}$ & $2.11\times 10^{9}$ & $3.55\times 10^{6}$ & $2.77\times 10^{4}$ & $1.30\times 10^{3}$ & 415 & - \\
False & 0.01 & $16$ & $1.62\times 10^{5}$ & 131.05 & 41 & $4.41\times 10^{4}$ & $5.17\times 10^{8}$ & $1.78\times 10^{6}$ & $6.29\times 10^{3}$ & $1.00\times 10^{4}$ & 729 & 3.12 \\
\end{tblr}

  \end{center}
  \caption{
  Resource requirements for the Toeplitz matrix use case.
    Number of phase factors characterises how many phase factors are required to approximate $A^{-1}$ at accuracy $\epsilon$, $\kappa$ is the subnormalisation of the matrix to be inverted.
    Oracle time is the time taken to implement an amplitude encoding oracle where Amplitude Estimation is True, and a single QSVT circuit otherwise.
    Oracle calls is given by Equation~\ref{eq:query_complexity} when Amplitude Estimation is True, and $\frac{1}{\epsilon^2}$ otherwise.
    Total time is oracle calls multiplied by oracle time.
    Percentage of amplitudes characterises the percentage of the basis vectors that can be measured using Amplitude Estimation in the time taken for the naive case.
}\label{tab:toeplitz_resource_requrements}
\end{table}
\end{turnpage}


\end{document}